\documentclass[twocolumn]{aastex631}

\usepackage{xcolor}
\usepackage{tabularx}
\usepackage{amsmath}

\begin{document}

\title{Searching for Quiescent Galaxies over $3 < z < 6$ in JWST Surveys Using Manifold Learning}

\author[0000-0002-6219-5558]{Alexander de la Vega}
\affiliation{Department of Physics and Astronomy, University of California, 900 University Ave, Riverside, CA 92521, USA}

\correspondingauthor{Alexander de la Vega}
\email{alexandd@ucr.edu}

\author{Mitchell D. Babcock}
\altaffiliation{Deceased March 2024.}
\affiliation{Department of Astronomy and Astrophysics, University of California, Santa Cruz, 1156 High Street, Santa Cruz, CA 95064, USA}
\affiliation{Department of Physics and Astronomy, University of California, 900 University Ave, Riverside, CA 92521, USA}

\author[0000-0001-5846-4404]{Bahram Mobasher}
\affiliation{Department of Physics and Astronomy, University of California, 900 University Ave, Riverside, CA 92521, USA}

\author{Dominik A. Riemann}
\affiliation{Department of Astronomy, University of Washington, Seattle, WA 98195, USA}
\affiliation{Department of Physics, University of Washington, Seattle, WA 98195, USA}
\affiliation{Department of Physics and Astronomy, University of California, 900 University Ave, Riverside, CA 92521, USA}

\author[0000-0003-3691-937X]{Nima Chartab}
\affiliation{IPAC, California Institute of Technology, Pasadena, CA 91125, USA}

\author[0000-0003-2226-5395]{Shoubaneh Hemmati}
\affiliation{IPAC, California Institute of Technology, Pasadena, CA 91125, USA}

\author[0000-0002-7530-8857]{Arianna S. Long}
\affiliation{Department of Astronomy, University of Washington, Seattle, WA 98195, USA}

\author[0009-0009-3048-9090]{Sogol Sanjaripour}
\affiliation{Department of Physics and Astronomy, University of California, 900 University Ave, Riverside, CA 92521, USA}



\begin{abstract}
Quiescent galaxies over $3<z<6$ are rare and puzzling. They formed and quenched within two billion years and simulations routinely struggle to predict their observed abundances. Developing a robust identification technique for these galaxies is crucial for constraining galaxy evolution models. Traditional rest-frame color-color selection techniques for quiescent galaxies are known to break down or require adjustments at $z\gtrsim3$. Recently, observed-frame color-color criteria have been established with JWST/NIRCam colors that efficiently pre-select high-redshift quiescent galaxies using only $\lesssim1\%$ of a given sample. In this work, the Uniform Manifold Approximation and Projection machine-learning technique is applied to pre-select quiescent galaxies over $3<z<6$ using observed NIRCam colors. From a parent sample of 43,926 galaxies in JADES, we ultimately find 44 quiescent candidates from a pool of $\approx2,300$ galaxies. This is about five times fewer galaxies than what would be pre-selected using color-color criteria. Two-thirds of these candidates can be pre-selected from a pool as small as 247, which is about twice as efficient as existing observed-frame color selection techniques. Nearly two-thirds of the candidates are new discoveries and include quiescent galaxies with mass-weighted ages as young as $\lesssim300$ Myr. We obtain number densities in agreement with the literature at $z<4$ and find generally higher abundances at $z>4$, although our measurements are consistent within errors. This technique may be applied to other JWST surveys. 
\end{abstract}



\section{Introduction} \label{sec:intro}

Determining the cause(s) behind the surprisingly high abundance of quiescent galaxies in the first few billion years after the Big Bang is an important open question in galaxy evolution \citep{Nayyeri14, Toft14, Glazebrook17, Schreiber18, Merlin19, Forrest20, Shahidi20, Valentino20}. Most cosmological models produce too few quiescent galaxies at $z>3$ when compared with observations, especially at high stellar masses, $M_{\star} \gtrsim 10^{10}M_{\odot}$ \citep{Steinhardt16, Schreiber18, Cecchi19, Merlin19, Lovell23, Lagos24, Vani25}. This indicates significant shortcomings in our understanding of the physical mechanisms needed to rapidly end the star-formation in these systems. 

To better understand how quiescent galaxies formed at $z\gtrsim3$, large samples of them must be obtained. However, these galaxies are very rare, with number densities of about $10^{-6} - 10^{-5}~\textrm{Mpc}^{-3}$ \citep{Nayyeri14, Girelli19, Santini19, Shahidi20, Valentino20, Valentino23, Carnall23_MNRAS, Long23, Long24, Alberts24, Baker24}. They have traditionally been identified using rest-frame color-color diagrams. The main aspects of these are the use of an optical color to straddle the Balmer break, which is sensitive to age, and a near-infrared (NIR) color to distinguish between dusty star-forming and quiescent galaxies. The most common is the $UVJ$ diagram, which works well at separating quiescent from star-forming galaxies at $z<2$, where a distinct bimodality in color-color space exists \citep[e.g.,][]{Williams09, Whitaker11, Brammer11, Belli19}. At higher redshifts, the bimodality is less clear as quiescent galaxies are younger (ages of $\lesssim500$ Myr) and thus have bluer colors and a significantly more efficient gas removal and star formation process at earlier epochs \citep{Straatman16, DEugenio20, Forrest20}. This is likely due to their quenching rapidly at early times \citep[e.g.,][]{Pacifici16, Carnall18}, which prevents them from aging enough to have colors as red as quiescent galaxies at lower redshifts \citep{Merlin19, Lovell23}. As a result, studies relying on the $UVJ$ diagram at high redshift report the need to adjust color-color selection regions \citep[e.g.,][]{Valentino23}, significant contamination rates (10-40\%) from dusty star-forming galaxies \citep[][]{Deshmukh18, Martis19}, and incompleteness \citep{Deshmukh18, Merlin18, Valentino20, Lovell23, Lustig23}. 

Alternative techniques have recently been developed to both combat the shortcomings of the $UVJ$ diagram and make use of the unparalleled wavelength coverage and sensitivity of the James Webb Space Telescope (JWST, \citealt{Gardner23}). \citet{AntwiDanso23_ugi} devised a rest-frame color-color diagram using synthetic colors in the $u, g, $ and $i$ bandpasses. This combination has been shown to reduce contamination from dusty star-forming and emission-line galaxies by about a factor of two. \citet{Long24} took a different approach and proposed an {\it observed}-frame color-color diagram using the JWST/NIRCam F150W--F277W and F277W--F444W colors. In this method, a subset of galaxies making up $\lesssim1\%$ of their sample was selected and and examined for the presence of high-redshift quiescent galaxies. In other words, their observed-frame color-color diagram efficiently captures high-redshift quiescent galaxies, which are selected along with non-quiescent galaxies that have similar colors and therefore increases the rate of contamination. 

Another way to identify high-redshift quiescent galaxies, as well as rare populations of galaxies in general, is through machine learning algorithms (see \citealt{BallBrunner10} and \citealt{HuertasCompany23} for reviews). These techniques have been used to search for various rare (sub-)populations of galaxies, such as mergers \citep{Ackermann18}, active galactic nuclei \citep{Faisst19, Sanjaripour24}, and quiescent galaxies at $z<2$ \citep{Steinhardt20}. 
\citet{Gould23} used the machine learning technique of Gaussian mixture models to assign probabilities that galaxies are quenched. This was done by examining the location of galaxies in three rest-frame colors: $NUV-U, U-V,$ and $V-J$.
A noteworthy subset of machine learning techniques that have been applied to studies of galaxies is that of manifold learning. In these techniques, which include Self-Organizing Maps \citep[SOMs;][]{Kohonen01} and Uniform Manifold Approximation and Projection \citep[UMAP;][]{McInnes18}, multi-dimensional datasets are mapped onto lower-dimensional representations while preserving information contained in the higher-dimensional data. These techniques have been used successfully to measure physical properties from colors for observed \citep[e.g.,][]{Geach12, Masters15, Hemmati19, Davidzon22} and simulated \citep{Davidzon19, Simet21, LaTorre24} galaxies and even estimate the fluxes of missing data \citep{Chartab23, LaTorre24}. 

In this work, we use the UMAP machine learning technique to pre-select quiescent galaxies over $3 < z < 6$ using JWST/NIRCam photometry. Our method is built on clustering together galaxies that have similar observed-frame colors to build a pool of candidates, which likely contains high-redshift quiescent galaxies, from a much larger sample. Quiescent candidates are found by fitting the spectral energy distributions (SEDs) of the galaxies that are pre-selected. Multiple wide and/or deep JWST surveys have already observed in the bandpasses used in this technique. Therefore, large samples of high-redshift quiescent galaxies may be obtained by applying our technique to other datasets.

This paper is organized as follows. The dataset and selection criteria of the observational sample used to identify high-redshift quiescent galaxies are defined in Section \ref{sec:data}. The seven observed-frame colors used to identify quiescent galaxies are presented in Section \ref{sec:training_colors}, along with the models used to train our machine learning algorithm. We describe and apply our technique to observations in Section \ref{sec:identify_qgs}. Two different ways of searching for quiescent galaxies are developed in Section \ref{sec:sample_qg}. A final sample of quiescent galaxies at $3<z<6$ is presented and our technique is compared with the observed color-color selection method developed by \citet{Long24}. Number densities are also measured. Our conclusions are summarized in Section \ref{sec:conclusions}. 

Throughout this work we use the AB photometric system \citep{OkeGunn83} and adopt the cosmological parameters measured by \citet{Planck16}. 

\section{Data and Selection of the Testing Set}
\label{sec:data}

Photometry and spectroscopic redshifts measured by the JWST Advanced Deep Extragalactic Survey (JADES, \citealt{Eisenstein23a, Eisenstein23b, Rieke23, DEugenio24}) survey and ancillary datasets are used. Details on these data are provided below. 

\subsection{Source Detection and Photometry}
\label{sec:photometry}
This work is based on NIRCam observations of the Great Observatories Origins Deep Survey (GOODS; \citealt{Giavalisco04}) North and South (hereafter, GOODS-N and GOODS-S) fields taken by JADES. The combined area of these fields is 124 arcmin$^2$. The GOODS-N field was observed in 11 NIRCam bandpasses (F090W, F115W, F150W, F182M, F200W, F210M, F277W, F335M, F356W, F410M, and F444W) and GOODS-S was observed in these and three additional bandpasses (F430M, F460M, and F480M). The medium-band filters F182M, F210M, F430M, F460M, and F480M were observed in GOODS-S by the JEMS program \citep{Williams23}. The 5$\sigma$ point-source depth measured in 0.3\arcsec \ circular apertures is $\sim29.5$ AB mag in each bandpass. 

Source detection in JADES and photometry in nine NIRCAM bandpasses and 14 ground- and space-based bandpasses are performed by \citet{Rieke23}. Briefly, a detection image is created by stacking the F277W, F335M, F356W, F410M, and F444W images and weighting them by their inverse variances. Next, various source detection routines in the {\tt python} module {\tt Photutils} \citep{Bradley23} are employed to detect and deblend sources and perform photometry. All NIRCam images are aligned to the same reference frame and drizzled to a common pixel scale of 0.03\arcsec \ per pixel. Photometry is performed on HST and NIRCam images convolved to the F444W resolution. See Section 4 by \citet{Rieke23} for more details. 

This work uses photometry in all NIRCam bandpasses above, where available, and is measured in Kron apertures with a Kron parameter of $k=2.5$. Photometry in the HST/Advanced Camera for Surveys F435W, F606W, and F775W bandpasses is also used in this work, which comes from the GOODS and Cosmic Assembly Near-infrared Deep Extragalactic Legacy Survey \citep[CANDELS;][]{Grogin11, Koekemoer11} surveys. HST photometry is performed by the JADES team in the same way as that for JWST bandpasses. 

Errors on photometry come from two sources: errors due to sky subtraction and a combination of source Poisson and instrumental errors. The first are measured in apertures of various sizes that are centered on empty regions of the sky (see Section 4.2 in \citealt{Rieke23}, also \citealt{Labbe05}; \citealt{Whitaker11}; and \citealt{Skelton14}). The second come from the ERR extension in the JADES images. Total errors on the JWST photometry are computed by adding these two sources of error in quadrature. For the HST images, the JADES team have measured errors due to sky subtraction and report the average values of the weight, or inverse variance, within the Kron apertures. The error due to the weight for each galaxy is computed by multiplying the average weight value by the area of the Kron aperture in pixels. Total errors on the HST photometry are calculated by adding these two errors in quadrature. A final error is added to all fluxes for each bandpass to account for zeropoint errors and mismatches in the templates between model and observed SEDs. It is equal to 5\% of the flux (see, e.g., Table 4 of \citealt{Dahlen13} and Table 11 of \citealt{Skelton14}). 

\subsection{Photometric and Spectroscopic Redshifts}
Photometric redshifts in JADES are estimated by \citet{Hainline24} using {\tt EAZY} \citep{Brammer08}. {\tt EAZY} obtains photometric redshifts by fitting non-negative linear combinations of templates to observed photometry in order to derive probability distribution functions of the redshift. Templates must be supplied by the user. \citet{Hainline24} assume modified versions of the original galaxy templates used by \citet{Brammer08} and develop seven new templates designed to fit the SEDs of high-redshift galaxies (see Section 3 by \citealt{Hainline24}). 

Spectroscopic redshifts come from JADES and the compilation over the CANDELS fields by \citet{Kodra23}. Descriptions of the selection of the targets, reduction and calibration of the data, measurements of the redshifts, and presentation of the JADES spectra are provided by \citet{Bunker24} and \citet{DEugenio24}. Descriptions of the provenances of the other spectroscopic redshifts and quality cuts applied can be found in \citet{Kodra23} and references therein. 

To assess the accuracy of the photometric redshifts measured by the JADES team, we compare them against the spectroscopic redshifts from all provenances above. There are 5,780 galaxies with spectroscopic redshifts spanning $0<z<13.35$. The accuracies of the photometric redshfits are quantified using the bias and normalized median absolute deviation (NMAD) scatter following \citet{Dahlen13}. We measure a bias of -0.003, such that the photometric redshifts very slightly underestimate the spectroscopic ones, and $\sigma_{\textrm{NMAD}} = 0.042$. 

\subsection{Selection of the Observational Sample}
\label{sec:select_test_set}

We begin with all sources from the JADES GOODS-N and JADES GOODS-S photometric catalogs. There is a total of 179,709 sources in both fields: 85,709 in GOODS-N and 94,000 in GOODS-S. We first select all sources with a brightness greater than 29th mag in all six bandpasses that make up the colors used to train the UMAP algorithm (F115W, F150W, F200W, F277W, F356W and F444W, see Section \ref{sec:selecting_colors}). This magnitude cut is used to achieve completeness and is brighter than the limiting magnitudes for all NIRCam filters in JADES \citep[Table 2 in][]{Rieke23}. This cut removes 115,223 objects with 64,486 remaining. Next, we remove all stars from the sample using the {\tt FLAG\_ST} flag. This cut removes 483 sources, leaving us with 64,003 galaxies. Finally, we remove all sources with adjacent bright stars or bright neighbors using the {\tt FLAG\_BS} and {\tt FLAG\_BN} flags. The former cut removes 1,327 galaxies and the latter removes 18,750. 

The final observational sample contains 43,926 sources: 21,554 in GOODS-N and 22,372 in GOODS-S. Galaxies in the final sample have a median signal-to-noise ratio (S/N) of about 9.0 in F115W and 13.7 in F444W. 

\section{Observed Colors and Selection of the Training Set}
\label{sec:training_colors}

\subsection{Selecting Observed Colors}
\label{sec:selecting_colors}

We use the observed colors instead of rest-frame ones. This is because selecting on rest-frame colors requires fitting SEDs for all galaxies in a sample, which may be costly and model depedendent. By selecting on observed colors first, most of the sample may be discarded, and only the subset of the sample that is likely to contain galaxies of interest may have its SEDs fit. 

Observed colors that span the rest-ultraviolet and rest-optical are used. In the absence of dust, quiescent galaxies can be separated from star-forming galaxies by selecting based on rest-optical colors that span the Balmer break at a rest-frame wavelength of $\sim4000~$\AA. This is because the strength of this spectral feature is correlated with the light-weighted age of a stellar population. However, quiescent galaxies at $z\gtrsim3$ tend to have young ages of about 500 Myr, which results in weaker Balmer breaks than those observed at lower redshifts \citep[e.g.,][]{DEugenio20, Forrest20, Stevans21}. This has motivated the use of rest-frame near-ultraviolet (NUV) colors, which are more sensitive to rapid shutdowns in star-formation than rest-frame optical colors \citep[][and references therein]{Gould23}. 

\begin{figure*}[t!]
    \includegraphics[width=\textwidth]{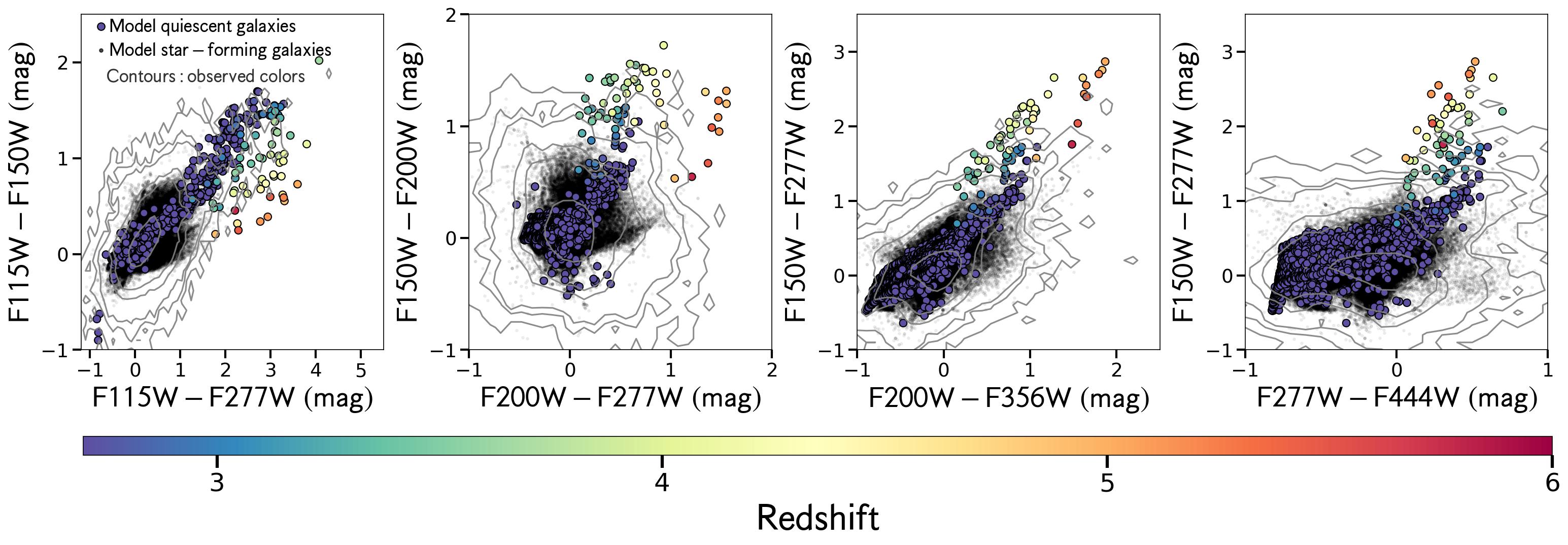}
    \caption{We search for quiescent galaxies by training a machine learning algorithm on seven observed-frame colors of model galaxies: F115W -- F150W, F115W -- F277W, F150W -- F200W, F150W -- F277W, F200W -- F277W, F200W -- F356W, and F277W -- F444W. These are chosen to separate high-redshift quiescent galaxies from others. Colors of galaxies in the observational sample (Section \ref{sec:select_test_set}) and model galaxies from the JAGUAR mock catalog \citep{Williams18} are plotted in four color-color diagrams. Observed colors are shown as gray contours indicating where the density is 1, 5, 10, 50, 100, and 500 galaxies. Model colors of quiescent galaxies are shown as dots colored by redshift and those of star-forming galaxies are shown as smaller black dots. Model galaxies are selected using the same magnitude cut as the testing set of observed galaxies (Section \ref{sec:photometry}). In the colors shown here, model quiescent galaxies at $z\geq3$ are separated from star-forming galaxies at all redshifts and queiscent galaxies at lower redshifts. The distributions of observed colors generally follow those of the models. The colors are based on the rest-frame $(ugi)_s$ \citep{AntwiDanso23_ugi} and $NUV-U, V-J$ color-color diagrams \citep{Gould23} and the observed-frame color-color diagram developed by \citet{Long24}, which is shown in the rightmost column. }
    \label{fig:color_color_diagrams}
\end{figure*}

We use six broadband JWST/NIRCam filters to devise seven colors that span the rest-frame NUV to NIR at $3<z<6$: F115W -- F150W, F115W -- F277W, F150W -- F200W, F150W -- F277W, F200W -- F277W, F200W -- F356W, and F277W -- F444W. These colors were chosen for two reasons. First, model high-redshift quiescent galaxies are relatively isolated from other models in these colors (Figure \ref{fig:color_color_diagrams} and Section \ref{sec:training_set}). Second, these colors span a wide wavelength range and contain enough information to separate model galaxies by redshift (middle panel of Figure \ref{fig:umap_ssfr_z_tauv} and Section \ref{sec:2d_viz}). 

\subsection{Selection of the Training Set from a Mock Catalog}
\label{sec:training_set}

To identify quiescent galaxies at high redshift, we first train a machine learning algorithm on model colors, which produces a two-dimensional visualization (details are provided in Section \ref{sec:2d_viz}). This visualization serves as a map with $x$ and $y$ coordinates; model galaxies with similar colors have similar coordinates. Then, we map the observed colors onto the visualization. Models are from the JAdes extraGalactic Ultradeep Artificial Realizations mock catalog \citep[JAGUAR;][]{Williams18}. JAGUAR is a phenomenological model that describe the evolution of galaxy number counts, morphologies, and SEDs over redshifts $0.2< z< 15$ and a wide range in stellar mass $\left(M_{\star} \geq 10^6 M_{\odot}\right)$. It provides mock photometry in more than a dozen NIRCam bandpasses and also includes a self-consistent treatment of stellar and photoionized gas emission and dust attenuation based on the {\sc beagle} SED-fitting tool \citep{Chevallard16}. For more details, see \citet{Williams18}. 

JAGUAR assigns mock galaxy SEDs by drawing from a parent sample of observed SEDs using multi-wavelength photometric catalogs from the 3D-HST survey \citep{Skelton14}. Quiescent galaxies are identified and separated based on their $UVJ$ colors and must have specific star-formation rates (sSFRs, or SFR divided by stellar mass) $<10^{-10}~\textrm{yr}^{-1}$. Therefore, the full diversity of $z > 3$ quiescent galaxies is not covered in this parent sample, particularly the bluer, younger populations. However, as shown in Section \ref{sec:sample_qg}, our method is successful in identifying many young ($< 300$ Myr) quiescent galaxies, thus we believe that the impact of a limited range in quiescent SEDs is negligible. We note that JAGUAR contains quiescent models up to $z\approx5.8$. 

Our training set comprises all mock galaxies with magnitudes brighter than 29 in all six of the following bandpasses: F115W, F150W, F200W, F277W, F356W, and F444W. The magnitude cut is the same as that applied to the observational sample (Section \ref{sec:photometry}). JAGUAR provides two mock catalogs, one for star-forming galaxies and another for quiescent ones. The same cut is applied to both, which results in a training set with 74,690 mock galaxies: 73,037 star-forming and 1,653 quiescent. 

The seven colors used to distinguish high-redshift quiescent galaxies from others are shown in four color-color diagrams in Figure \ref{fig:color_color_diagrams}. Each diagram contains observed and mock colors. The former are shown as light gray contours. The latter are shown in different ways for quiescent and star-forming models. High-redshift, i.e. $z\geq3$, quiescent galaxies are colored by their redshift. Lower-redshift quiescent galaxies, shown as dark blue dots, are offset from their higher-redshift counterparts and overlap with star-forming galaxies, which are plotted as small black points. In each diagram, the distribution of observed colors overlaps well with the models and high-redshift quiescent galaxies are outliers in at least one color. 
This suggests that high-redshift quiescent galaxies in JAGUAR are relatively isolated compared with the rest of the simulated galaxies in the multidimensional color space. This will make it easier to identify high-redshift quiescent galaxies when applying our machine learning algorithm (Section \ref{sec:identify_qgs}). 

\section{Identifying High-Redshift Quiescent Galaxies in JADES}
\label{sec:identify_qgs}

In this section, we search for high-redshift quiescent galaxies in JADES. We provide a definition of quiescence that is used to select model quiescent and observed galaxies in Section \ref{sec:def_qg}. Then, we explain how the UMAP machine learning algorithm is trained on model galaxy colors in JAGUAR in Section \ref{sec:2d_viz}. The resulting two-dimensional visualization is described. The observed colors of galaxies in JADES are mapped onto the visualization in Section \ref{sec:mapping}. We show in Section \ref{sec:baker24} that recent JWST spectroscopic observations confirm that our technique can quickly identify high-redshift quiescent galaxies. Fits to the SEDs of galaxies in the observational sample are described in Section \ref{sec:bagpipes}. 

\begin{figure*}[t!]
    \includegraphics[width=\textwidth]{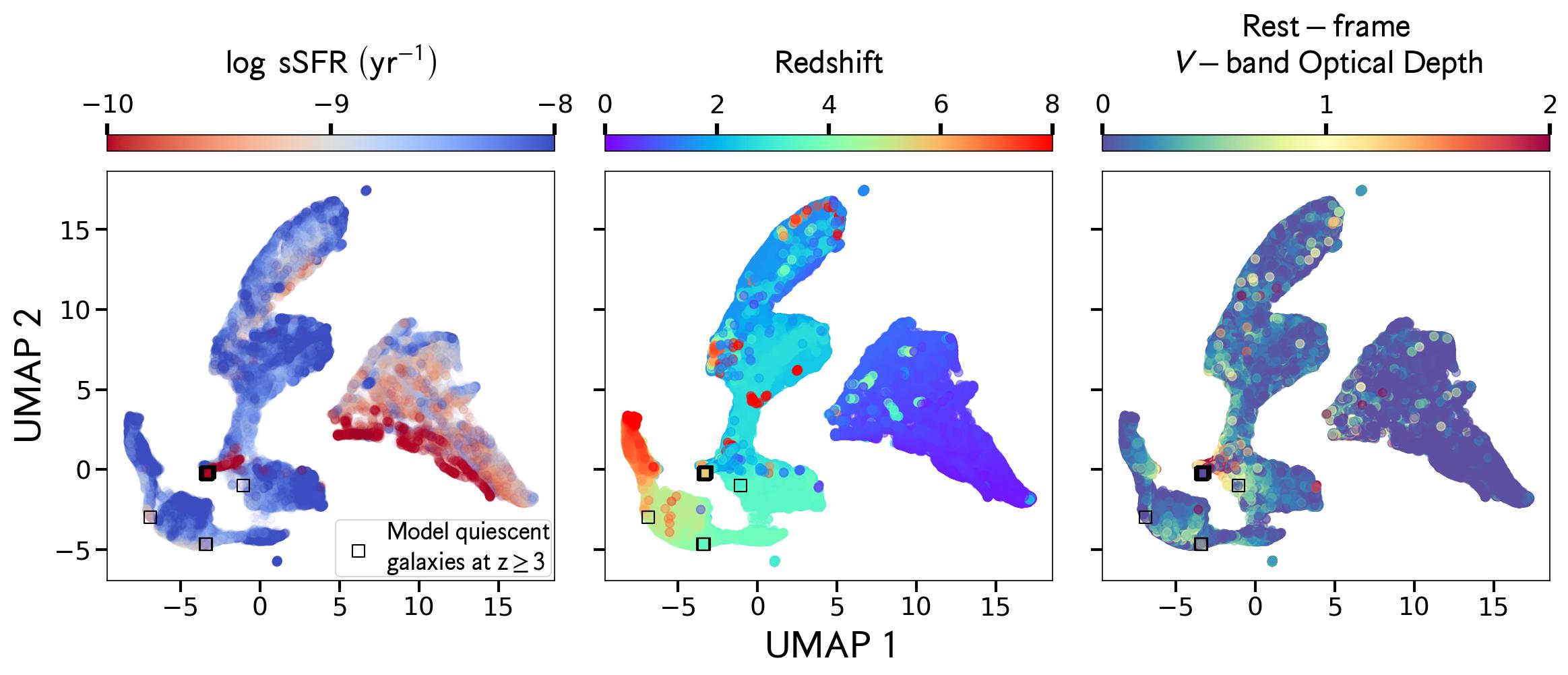}
    \caption{This figure shows a two-dimensional visualization of the JAGUAR models using the UMAP technique. It is trained on the seven NIRCam colors shown in Figure \ref{fig:color_color_diagrams}. From left to right, models are colored by the specific star-formation rate, redshift, and rest-frame $V$-band optical depth. There are clear qualitative trends in the visualization. Models are mapped to three large groupings. The highest-redshift models are on the left, and the lowest-redshift ones are on the right. Model sSFRs are generally high and dust values are generally low. Rare models, such as low-sSFR and dusty models, are located at the edges of the groupings. The 62 model quiescent galaxies at $z\geq 3$ (Section \ref{sec:def_qg}) are located in the lower left of the visualization. They are indicated by the empty black squares. Of these, 55 are clustered tightly at coordinates (UMAP 1, UMAP 2) $\approx (-3.3, -0.2)$. A second, smaller group of five models is located at approximately ($-3.5, -5$). }
    \label{fig:umap_ssfr_z_tauv}
\end{figure*}

\subsection{Defining Quiescence}
\label{sec:def_qg}
A galaxy is determined to be quiescent given the value of its sSFR with respect to the age of the Universe at the redshift of the galaxy:
$$\textrm{sSFR}_{100} \leq 0.2 / t_{U},$$
where $\textrm{sSFR}_{100}$ is the sSFR averaged over the last 100 Myr since the time of observation and $t_U$ is the age of the Universe at the redshift of the galaxy in yr. This cut in sSFR has been used in previous studies \citep[e.g.,][]{Pacifici16, Carnall2020, Alberts24, Baker24} and has been shown to be consistent with a color cut in the $UVJ$ diagram \citep{Gallazzi14}. 

This criterion is used to select both model and observed quiescent galaxies. There are 62 model galaxies in the training set selected from JAGUAR that satisfy this criterion at $z\geq3$. The SFRs and stellar masses for the observational sample come from SED fitting (Section \ref{sec:bagpipes}). We determine whether an observed galaxy is quiescent using the median of the posterior probability distribution function on the sSFR. 

\begin{figure*}[t!]
    \includegraphics[width=\textwidth]{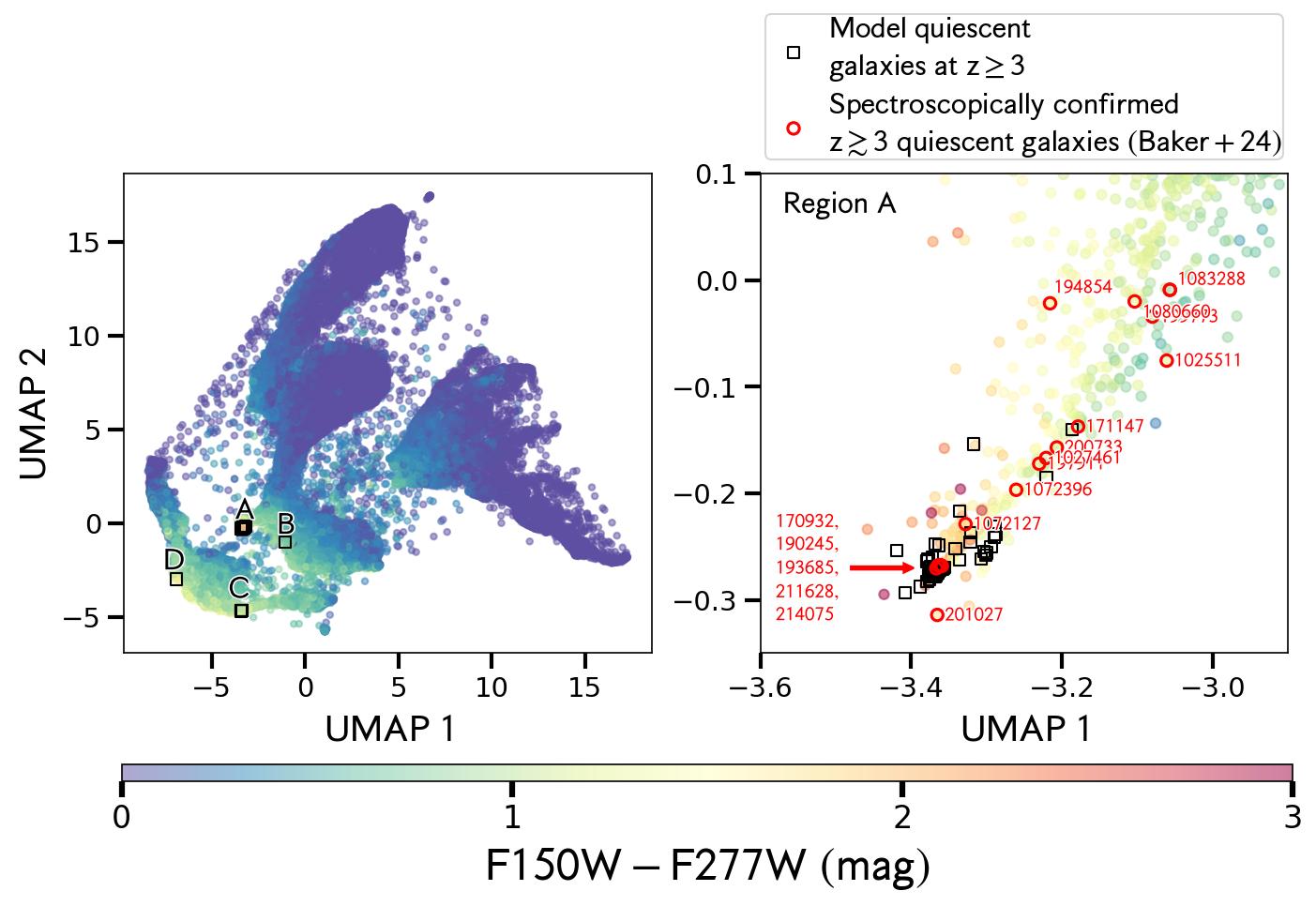}
    \caption{This figure shows the mapping of observed galaxy colors in JADES onto the UMAP visualization on the left. Galaxies are shown as dots colored by F150W-F277W color. The observed colors are largely mapped onto the same regions as the JAGUAR models, though some are mapped onto regions where there are no models. Quiescent models are grouped into four regions labeled A through D. The right panel is centered on region A, which contains the bulk (55/62) of the $z\geq3$ quiescent galaxy models. They are plotted as empty black squares in both panels. Spectroscopically confirmed massive quiescent galaxies at $z\gtrsim3$ from \citet{Baker24} are shown as empty red circles. The sample in this paper contains 17 of the 18 galaxies studied in that work, all of which lie close to the $z\geq3$ quiescent models. This suggests that our technique efficiently captures high-redshift quiescent galaxies. The galaxies with spectra are labeled with their IDs listed in the JADES GOODS-N and GOODS-S photometric catalogs. Galaxy ID 194854 is also examined by \citet{Carnall23_Nature}. }
    \label{fig:umap_jades_baker}
\end{figure*}

\subsection{Two-Dimensional Visualization of the JAGUAR Models Using UMAP}
\label{sec:2d_viz}

The goal of our technique is to pre-select a subset of galaxies in our observational sample that is likely to contain high-redshift quiescent galaxies. Based on similar observed-frame color-selection methods \citep[e.g.,][]{Long24}, it is anticipated that any subset that is selected will contain many contaminants, due to the similarity in color between high-redshift quiescent galaxies and high-redshift dusty and emission-line star-forming galaxies \citep[e.g.,][]{Schreiber18, AntwiDanso23_ugi, McKinney23, Long24}. Despite the high contamination rate, the benefit of performing such a pre-selection is clear: instead fitting $\sim40,000$ galaxy SEDs (i.e., the size of the observational sample) to find a few dozen quiescent galaxies over $3<z<6$, only a few hundred or a few thousand would need to be fit. 

The pre-selection is based on principles of dimensionality reduction. The training set comprises a seven-dimensional parameter space, with one dimension corresponding to one color (Section \ref{sec:training_colors}). Galaxies that have similar values for all seven colors will have similar positions in this seven-dimensional space. It is therefore expected that galaxies with similar positions in this multi-dimenional space have similar physical properties, e.g., redshift, sSFR, and dust. However, these seven dimensions cannot be presented directly, so we appeal to techniques that can reduce the number of dimensions to just two while preserving the information contained in higher dimensions. 

In this paper, we use the UMAP technique \citep{McInnes18} to perform dimensionality reduction, create a visualization of the JAGUAR models, and map the observational sample onto the visualization to search for high-redshift quiescent galaxies. 
UMAP is a nonlinear technique that estimates the topology of a high-dimensional dataset and creates a lower-dimensional representation that preserves structural information on local and global scales. Briefly, it first constructs a graph of nearest neighbors in high-dimensional space. Each data point is connected to its $n$-nearest neighbors, where $n$ is set by the user. Every connection between any set of two points is weighted by the probability that these points are connected, which depends on the distance between them. This distance depends on the similarity of their properties. The user must also specify a minimum distance between points. The high-dimensional space is then projected to lower-dimensional embeddings. The structure of the graph is preserved as much as possible. 

\begin{figure*}[t!]
    \includegraphics[width=0.4\textwidth]{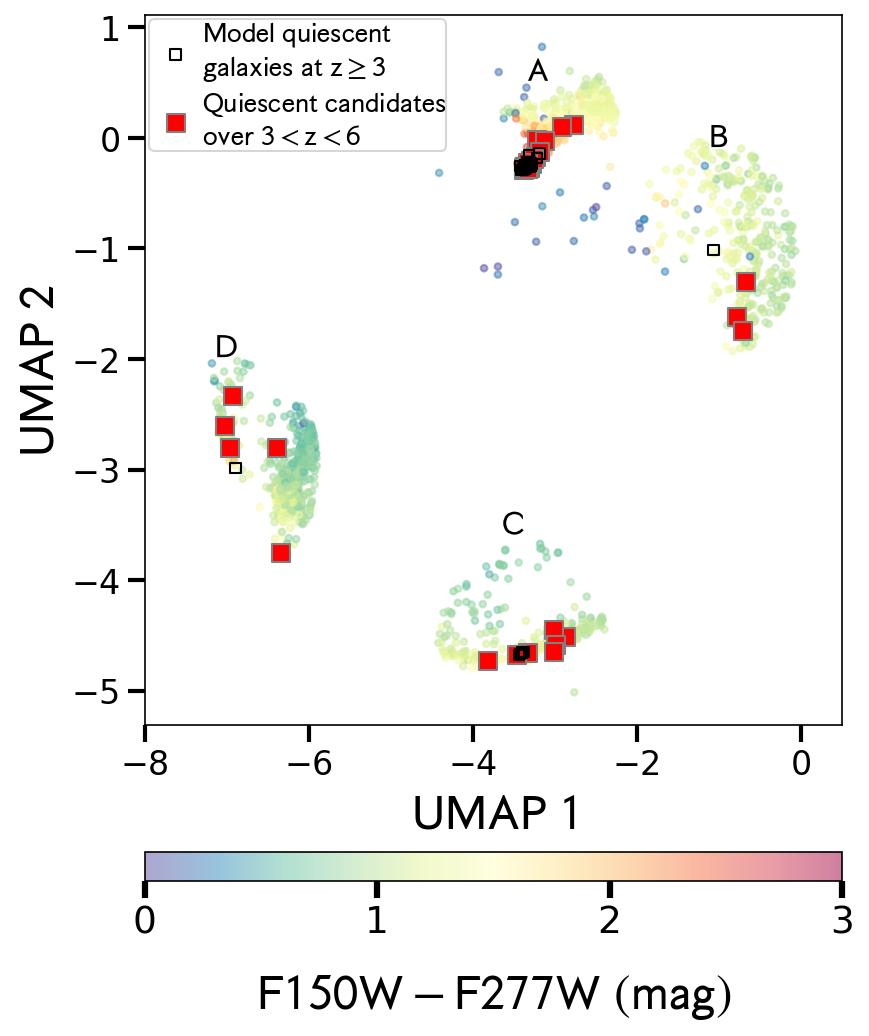}
    \includegraphics[width=0.6\textwidth]{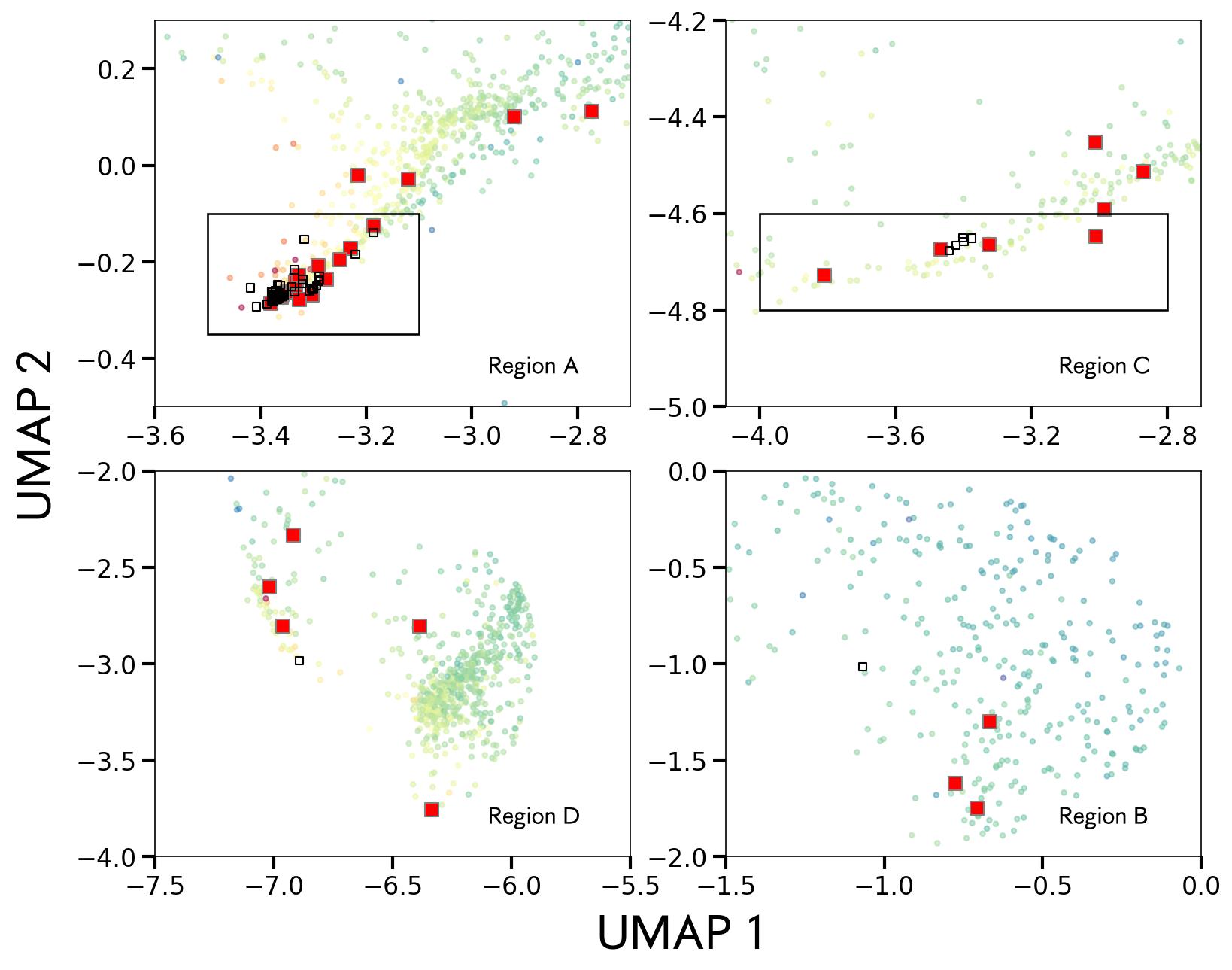}
    \caption{Quiescent galaxy candidates are identified by fitting the SEDs of galaxies that are near quiescent galaxy models. This is done in two techniques, shown here. In the first, for each model, shown as an empty square above, all galaxies that lie within a radius of value 1 in the UMAP parameter space are considered (Section \ref{sec:preselect_radial}). This pre-selection produces a sample of 2,282 galaxies, which are shown as dots colored by their F150W-F277W color on the left. Of these, we find 44 quiescent galaxy candidates at $3<z<6$, which are plotted as filled red squares. On the right, we zoom in to the four regions (labeled A through D) where the model quiescent galaxies are mapped onto the UMAP. We find that 29 out of 44 candidates can be pre-selected using two simple rectangular cuts in the projected space in Regions A and C (Section \ref{sec:preselect_rectangle}). These 29 include the oldest quiescent galaxies in our sample, and almost all lie within the color-color diagram proposed by \citet{Long24}. However, about half as many galaxies are pre-selected in the proposed rectangular regions in UMAP space than those chosen using the \citet{Long24} diagram (Figure \ref{fig:long24_diagram} and Section \ref{sec:efficiency}). }
    \label{fig:umap_radial_selection}
\end{figure*}

The UMAP technique has been shown to outperform other dimensionality reduction techniques, such as $t$-Distributed Stochastic Neighbor Embedding \citep[$t$-SNE;][]{vdM_Hinton08} and principal component analysis (see Sec. 4 in \citealt{Chartab23} and Sec. 4.4 in \citealt{Cook24}). 

We use the {\tt python} UMAP library\footnote{\href{https://github.com/lmcinnes/umap}{https://github.com/lmcinnes/umap}} to map the seven-dimensional color space of the JAGUAR models onto two dimensions. We set $n$ to 100 and the minimum distance to 0.01. These values were chosen by sweeping through a wide range of values for each parameter and examining how clustered quiescent galaxy models at $3<z<6$ are within the visualization. The high-redshift quiescent models are maximally clustered assuming these values. We note that the size of the visualization is determined while training the algorithm on the input data. It depends on the choice of the number of nearest neighbors; a smaller number results in a more compact visualization and vice versa. Smaller values of $n$ and therefore more compact visualizations result in looser clusterings of model high-redshift quiescent galaxies. 

In Figure \ref{fig:umap_ssfr_z_tauv}, we show the visualization created by UMAP that was trained on the JAGUAR models. Individual model galaxies are colored by the log sSFR, redshift, and rest-frame $V$-band optical depth from left to right. 
There are clear qualitative trends in the visualization. Models are mapped to three large groupings, which are structured according to redshift such that the highest-redshift models are on the left and the lowest-redshift ones are on the right. The model sSFR is high for nearly all high-redshift models, except for a small region in the lower left corner. This region contains the high-redshift quiescent models, which are shown as empty squares. Dusty models are found at the edges of each grouping. 

Out of the 62 high-redshift quiescent models, 55 are clustered tightly at coordinates (UMAP 1, UMAP 2) $\approx (-3.3, -0.2)$, which we refer to as ``Region A'' throughout the rest of the paper. A second, smaller group of five models is located at approximately ($-3.5, -5$) (``Region C''). The remaining two models are found between these groups, with one to the left (``Region D'') and the other to the right (``Region B''). This tight grouping of the models is encouraging, as it suggests that observed quiescent galaxies at $z\geq3$ may be found near the models, assuming the model colors are representative. We examine this in greater detail in Section \ref{sec:baker24}. 

\subsection{Mapping the Observations onto the Visualization}
\label{sec:mapping}

We now map the observational sample (Section \ref{sec:select_test_set}) onto the visualization created by the UMAP technique. This is done using the {\tt transform} routine in the UMAP {\tt python} library. This task takes a new dataset (in our case, the observational sample) and maps it onto a trained model by searching for the nearest neighbors. The mapping of the observational sample onto the UMAP visualization is shown in the left panel of Figure \ref{fig:umap_jades_baker}. Observed galaxies, shown as points colored by their F150W-F277W colors, largely fall into the same groupings as the models shown in Figure \ref{fig:umap_ssfr_z_tauv}. Some galaxies lie outside these groupings, which suggests they may have colors that are not reproduced by the models, have low S/N measurements, and/or are affected by artifacts. Regions A through D, which are populated by the quiescent models, are labeled in the left panel. 

\subsubsection{Validation Using Spectroscopic Observations}
\label{sec:baker24}

A subset of the observational sample has been observed by JWST/NIRSpec and was recently examined by \citet{Baker24}. Their sample of 18 galaxies consists of massive $(M_{\star} \gtrsim 10^{10}M_{\odot})$, spectroscopically confirmed quiescent galaxies over $2<z<5$. There are 17 galaxies in common between the observational sample of this work and the sample studied by \citet{Baker24}. One (JADES GOODS-N 1088564) is not in the sample examined in this work because it was not observed in the F277W and F356W bandpasses, which are used to train the UMAP algorithm. The 17 in common are plotted in the right panel of Figure \ref{fig:umap_jades_baker} and indicated by the empty red circles. All 17 lie in Region A and within a radius of 0.5 in the UMAP parameter space of the nearest $z\geq3$ quiescent models. Five galaxies coincide exactly with the bulk of the models. The close proximity between the models and these spectroscopically confirmed galaxies suggests that simulated galaxies in JAGUAR and the selection method of observed high-redshift quiescent galaxies by \citet{Baker24} are in agreement.

\begin{deluxetable*}{c c c c c c c c c}[t!]
\tablecaption{Final Sample of Quiescent Galaxy Candidates over $3<z<6$ in JADES\label{tab:qgs}}

\tablewidth{\textwidth}
\tablehead{ID & R.A. & Decl. & Redshift\tablenotemark{a} & $\log\left(M_{\star}\right)$ & $\log~\textrm{sSFR}_{100}$ & Mass-weighted Age & $A_V$ & References\tablenotemark{b} \\
& (deg) & (deg) & & $\left(M_{\odot}\right)$ & $\left(\textrm{yr}^{-1}\right)$ & (Gyr) & (mag) &}
\startdata
4348 & 53.06753379 & -27.90466113 & $3.460_{-0.193}^{+0.289}$ & $9.35_{-0.02}^{+0.02}$ & $-10.30_{-2.26}^{+1.19}$ & $0.13_{-0.01}^{+0.01}$ & $0.06_{-0.04}^{+0.06}$ & $\ldots$ \\
13124 & 53.11199689 & -27.89094615 & $3.410_{-0.591}^{+0.045}$ & $8.59_{-0.06}^{+0.05}$ & $-19.65_{-23.40}^{+8.08}$ & $0.40_{-0.08}^{+0.10}$ & $0.04_{-0.03}^{+0.07}$ & $\ldots$ \\
16170 & 53.08361282 & -27.88758582 & $3.390_{-0.098}^{+0.164}$ & $10.59_{-0.02}^{+0.02}$ & $-22.67_{-33.31}^{+9.63}$ & $0.64_{-0.14}^{+0.33}$ & $0.44_{-0.18}^{+0.16}$ & M19, C20, S20, A24 \\
37784 & 53.02296579 & -27.87239191 & $3.340_{-0.274}^{+0.420}$ & $8.89_{-0.08}^{+0.06}$ & $-13.02_{-14.59}^{+3.30}$ & $0.41_{-0.12}^{+0.15}$ & $0.13_{-0.09}^{+0.15}$ & $\ldots$ \\
40382 & 53.10151396 & -27.86995877 & $3.100_{-0.094}^{+0.346}$ & $10.73_{-0.05}^{+0.06}$ & $-56.63_{-164.85}^{+41.02}$ & $1.65_{-0.24}^{+0.19}$ & $1.23_{-0.21}^{+0.30}$ & $\ldots$ \\
73338 & 53.04274193 & -27.84866028 & $4.230_{-1.134}^{+0.262}$ & $8.81_{-0.08}^{+0.06}$ & $-13.60_{-17.23}^{+4.47}$ & $0.26_{-0.08}^{+0.06}$ & $0.12_{-0.08}^{+0.23}$ & $\ldots$ \\
104855 & 53.11667739 & -27.81093111 & $4.863_{-0.005}^{+0.005}$ & $8.52_{-0.05}^{+0.06}$ & $-15.39_{-14.53}^{+4.97}$ & $0.20_{-0.04}^{+0.08}$ & $0.07_{-0.05}^{+0.10}$ & $\ldots$ \\
163086 & 53.03776939 & -27.89074109 & $3.420_{-0.061}^{+0.163}$ & $10.66_{-0.02}^{+0.02}$ & $-19.56_{-27.11}^{+7.25}$ & $0.49_{-0.07}^{+0.16}$ & $0.30_{-0.15}^{+0.15}$ & $\ldots$ \\
163121 & 53.03492023 & -27.89089165 & $4.110_{-0.467}^{+0.324}$ & $8.80_{-0.07}^{+0.07}$ & $-11.80_{-11.38}^{+2.71}$ & $0.28_{-0.08}^{+0.17}$ & $0.13_{-0.09}^{+0.19}$ & $\ldots$ \\
163799 & 53.07379830 & -27.88919881 & $3.120_{-0.147}^{+0.270}$ & $9.11_{-0.02}^{+0.02}$ & $-13.37_{-5.82}^{+2.60}$ & $0.20_{-0.02}^{+0.01}$ & $0.07_{-0.05}^{+0.07}$ & $\ldots$ \\
170932 & 53.06226877 & -27.87504657 & $4.230_{-0.329}^{+0.124}$ & $10.36_{-0.04}^{+0.04}$ & $-10.01_{-2.00}^{+0.55}$ & $0.77_{-0.16}^{+0.14}$ & $0.61_{-0.27}^{+0.25}$ & M19, S20, A24, B24 \\
170972 & 53.13934870 & -27.87454432 & $3.450_{-0.168}^{+0.234}$ & $10.96_{-0.03}^{+0.03}$ & $-67.93_{-228.42}^{+47.72}$ & $1.60_{-0.11}^{+0.08}$ & $0.41_{-0.10}^{+0.12}$ & M19, C20 \\
172799 & 53.04750930 & -27.87050325 & $3.910_{-0.547}^{+0.075}$ & $11.09_{-0.04}^{+0.03}$ & $-10.29_{-4.06}^{+0.54}$ & $0.90_{-0.11}^{+0.19}$ & $0.86_{-0.26}^{+0.29}$ & A24 \\
174098 & 53.07586924 & -27.86885253 & $3.690_{-0.272}^{+0.107}$ & $9.09_{-0.04}^{+0.04}$ & $-11.29_{-5.18}^{+1.55}$ & $0.17_{-0.02}^{+0.02}$ & $0.05_{-0.03}^{+0.07}$ & $\ldots$ \\
174444 & 53.07867751 & -27.86834393 & $3.900_{-0.400}^{+0.228}$ & $8.92_{-0.04}^{+0.04}$ & $-13.49_{-10.60}^{+3.60}$ & $0.24_{-0.04}^{+0.06}$ & $0.09_{-0.07}^{+0.11}$ & $\ldots$ \\
174451 & 53.13802797 & -27.86829312 & $3.590_{-0.216}^{+0.137}$ & $10.55_{-0.03}^{+0.04}$ & $-42.65_{-108.40}^{+27.84}$ & $1.17_{-0.20}^{+0.21}$ & $0.34_{-0.19}^{+0.23}$ & C20 \\
175044 & 53.08283410 & -27.86618550 & $3.180_{-0.204}^{+0.370}$ & $10.47_{-0.03}^{+0.03}$ & $-10.24_{-6.16}^{+0.38}$ & $1.16_{-0.23}^{+0.20}$ & $0.76_{-0.22}^{+0.25}$ & $\ldots$ \\
182930 & 53.16152448 & -27.85607017 & $3.040_{-0.299}^{+0.048}$ & $10.42_{-0.04}^{+0.03}$ & $-15.66_{-26.66}^{+4.99}$ & $0.79_{-0.24}^{+0.12}$ & $0.09_{-0.06}^{+0.12}$ & $\ldots$ \\
184342 & 53.09163722 & -27.85340897 & $4.500_{-0.296}^{+0.089}$ & $11.63_{-0.04}^{+0.05}$ & $-23.40_{-40.41}^{+11.11}$ & $0.83_{-0.05}^{+0.03}$ & $0.84_{-0.13}^{+0.12}$ & $\ldots$ \\
185410 & 53.12440316 & -27.85169544 & $3.700_{-0.005}^{+0.005}$ & $10.77_{-0.04}^{+0.04}$ & $-13.55_{-10.23}^{+3.32}$ & $0.23_{-0.04}^{+0.07}$ & $0.38_{-0.12}^{+0.10}$ & $\ldots$ \\
189714 & 53.11350023 & -27.84131682 & $3.110_{-0.054}^{+0.170}$ & $9.76_{-0.04}^{+0.04}$ & $-11.36_{-5.49}^{+1.92}$ & $0.18_{-0.03}^{+0.05}$ & $0.46_{-0.12}^{+0.11}$ & $\ldots$ \\
190245 & 53.07872145 & -27.83961390 & $3.310_{-0.119}^{+0.071}$ & $10.21_{-0.02}^{+0.02}$ & $-25.77_{-50.85}^{+12.01}$ & $0.78_{-0.19}^{+0.27}$ & $0.38_{-0.18}^{+0.19}$ & M19, C20, S20, B24 \\
190674 & 53.07969892 & -27.83820594 & $3.400_{-0.087}^{+0.124}$ & $10.09_{-0.02}^{+0.02}$ & $-23.14_{-28.03}^{+10.80}$ & $0.51_{-0.10}^{+0.14}$ & $0.18_{-0.11}^{+0.14}$ & M19, C20 \\
193685 & 53.08188465 & -27.82880170 & $4.170_{-0.074}^{+0.079}$ & $10.33_{-0.03}^{+0.03}$ & $-10.95_{-10.19}^{+0.95}$ & $0.72_{-0.16}^{+0.09}$ & $0.45_{-0.17}^{+0.14}$ & M19, C20, S20, B24 \\
194854 & 53.10821066 & -27.82518639 & $4.570_{-0.069}^{+0.129}$ & $10.60_{-0.02}^{+0.02}$ & $-21.78_{-25.84}^{+8.90}$ & $0.31_{-0.02}^{+0.02}$ & $0.02_{-0.01}^{+0.04}$ & M19, C20, S20, B24 \\
197911 & 53.16531424 & -27.81413972 & $3.063_{-0.005}^{+0.005}$ & $11.17_{-0.04}^{+0.02}$ & $-12.56_{-4.01}^{+1.58}$ & $1.02_{-0.40}^{+0.07}$ & $0.26_{-0.11}^{+0.28}$ & M19 \\
198459 & 53.11912228 & -27.81403895 & $3.592_{-0.005}^{+0.005}$ & $10.50_{-0.04}^{+0.03}$ & $-13.93_{-18.22}^{+4.11}$ & $0.86_{-0.25}^{+0.14}$ & $0.95_{-0.22}^{+0.25}$ & R24 \\
199534 & 53.10410922 & -27.81022801 & $3.250_{-0.190}^{+0.375}$ & $9.61_{-0.02}^{+0.02}$ & $-29.31_{-45.85}^{+14.56}$ & $0.65_{-0.10}^{+0.21}$ & $0.05_{-0.04}^{+0.08}$ & $\ldots$ \\
214075 & 53.19691235 & -27.76052946 & $3.611_{-0.005}^{+0.005}$ & $10.69_{-0.02}^{+0.02}$ & $-34.91_{-51.31}^{+18.77}$ & $0.65_{-0.07}^{+0.09}$ & $0.43_{-0.10}^{+0.11}$ & M19, C20, S20, B24, R24 \\
214953 & 53.18123930 & -27.75650339 & $3.340_{-0.061}^{+0.086}$ & $10.59_{-0.02}^{+0.02}$ & $-23.49_{-35.52}^{+10.64}$ & $0.53_{-0.09}^{+0.14}$ & $0.60_{-0.16}^{+0.15}$ & M19, C20, S20, R24 \\
1000567 & 189.04895588 & 62.22093834 & $3.330_{-0.283}^{+0.395}$ & $8.72_{-0.09}^{+0.07}$ & $-10.52_{-14.82}^{+1.47}$ & $0.49_{-0.14}^{+0.30}$ & $0.19_{-0.12}^{+0.17}$ & $\ldots$ \\
1006121 & 189.03041796 & 62.24350022 & $4.840_{-0.644}^{+0.373}$ & $8.78_{-0.12}^{+0.12}$ & $-11.40_{-21.82}^{+2.41}$ & $0.45_{-0.17}^{+0.21}$ & $0.27_{-0.19}^{+0.30}$ & $\ldots$ \\
1016029 & 189.04781173 & 62.27850467 & $3.750_{-0.340}^{+0.029}$ & $10.59_{-0.03}^{+0.03}$ & $-35.67_{-88.05}^{+19.59}$ & $0.87_{-0.17}^{+0.22}$ & $0.46_{-0.22}^{+0.16}$ & $\ldots$ \\
1024921 & 189.17537421 & 62.22539286 & $4.424_{-0.005}^{+0.005}$ & $11.10_{-0.01}^{+0.01}$ & $-10.98_{-0.25}^{+0.15}$ & $0.79_{-0.04}^{+0.03}$ & $0.01_{-0.01}^{+0.02}$ & $\ldots$ \\
1025958 & 189.13792422 & 62.23601218 & $5.450_{-0.281}^{+0.147}$ & $9.86_{-0.06}^{+0.07}$ & $-10.53_{-8.91}^{+1.89}$ & $0.17_{-0.06}^{+0.14}$ & $0.66_{-0.39}^{+0.23}$ & $\ldots$ \\
1027147 & 189.07979795 & 62.24488130 & $3.020_{-0.083}^{+0.053}$ & $11.06_{-0.03}^{+0.02}$ & $-11.23_{-7.34}^{+0.99}$ & $1.03_{-0.44}^{+0.09}$ & $0.33_{-0.16}^{+0.24}$ & M19 \\
1027900 & 189.21331386 & 62.24985083 & $4.430_{-0.021}^{+0.123}$ & $9.85_{-0.09}^{+0.05}$ & $-9.88_{-5.32}^{+1.41}$ & $0.18_{-0.04}^{+0.04}$ & $0.17_{-0.12}^{+0.39}$ & $\ldots$ \\
1029481 & 189.06519419 & 62.26268910 & $5.520_{-0.114}^{+0.270}$ & $9.45_{-0.15}^{+0.16}$ & $-31.01_{-146.86}^{+17.96}$ & $0.71_{-0.25}^{+0.17}$ & $1.11_{-0.43}^{+0.48}$ & $\ldots$ \\
1034856 & 189.15578466 & 62.29852430 & $3.090_{-0.070}^{+0.089}$ & $10.27_{-0.04}^{+0.03}$ & $-15.55_{-16.55}^{+4.44}$ & $0.51_{-0.14}^{+0.35}$ & $0.28_{-0.17}^{+0.18}$ & $\ldots$ \\
1068658 & 189.21565373 & 62.14430461 & $3.880_{-0.058}^{+0.220}$ & $9.78_{-0.04}^{+0.04}$ & $-23.73_{-42.46}^{+10.78}$ & $0.68_{-0.20}^{+0.18}$ & $0.18_{-0.12}^{+0.18}$ & $\ldots$ \\
1069454 & 189.20977405 & 62.15168358 & $4.330_{-0.042}^{+0.014}$ & $9.41_{-0.02}^{+0.02}$ & $-24.00_{-29.22}^{+10.75}$ & $0.33_{-0.03}^{+0.03}$ & $0.01_{-0.01}^{+0.02}$ & $\ldots$ \\
1072127 & 189.26571758 & 62.16839324 & $4.129_{-0.005}^{+0.005}$ & $10.39_{-0.04}^{+0.04}$ & $-11.93_{-10.96}^{+2.22}$ & $0.43_{-0.09}^{+0.23}$ & $0.28_{-0.15}^{+0.17}$ & M19, S20, B24 \\
1076320 & 189.22145513 & 62.19240272 & $3.160_{-0.064}^{+0.088}$ & $10.21_{-0.03}^{+0.02}$ & $-27.79_{-54.29}^{+13.87}$ & $0.87_{-0.22}^{+0.31}$ & $0.36_{-0.22}^{+0.22}$ & S20 \\
1081823 & 189.23466913 & 62.22270467 & $3.260_{-0.223}^{+0.084}$ & $10.55_{-0.04}^{+0.04}$ & $-44.81_{-116.26}^{+28.04}$ & $1.52_{-0.33}^{+0.16}$ & $0.31_{-0.22}^{+0.16}$ & M19 \\
\hline
\enddata

\tablecomments{Medians of the stellar mass, sSFR, mass-weighted age, and $A_V$ posterior distributions are reported. Quantities in the subscript denote the differences between the 16th percentiles and the medians, and quantities in the superscript denote the differences between the 84th percentiles and the median. \\
$^a$ Spectroscopic redshifts have assumed uncertainties of $\pm0.005$. All other redshifts are photometric. \\
$^b$ M19 corresponds to \citet{Merlin19}; C20 corresponds to \citet{Carnall2020}; S20 corresponds to \citet{Shahidi20}; A24 corresponds to \citet{Alberts24}; B24 corresponds to \citet{Baker24}; and R24 corresponds to \citet{Russell24}. Ellipses ($\ldots$) indicate no match to a previous study was found.}

\end{deluxetable*}

\begin{figure*}[h]
    \includegraphics[width=\textwidth]{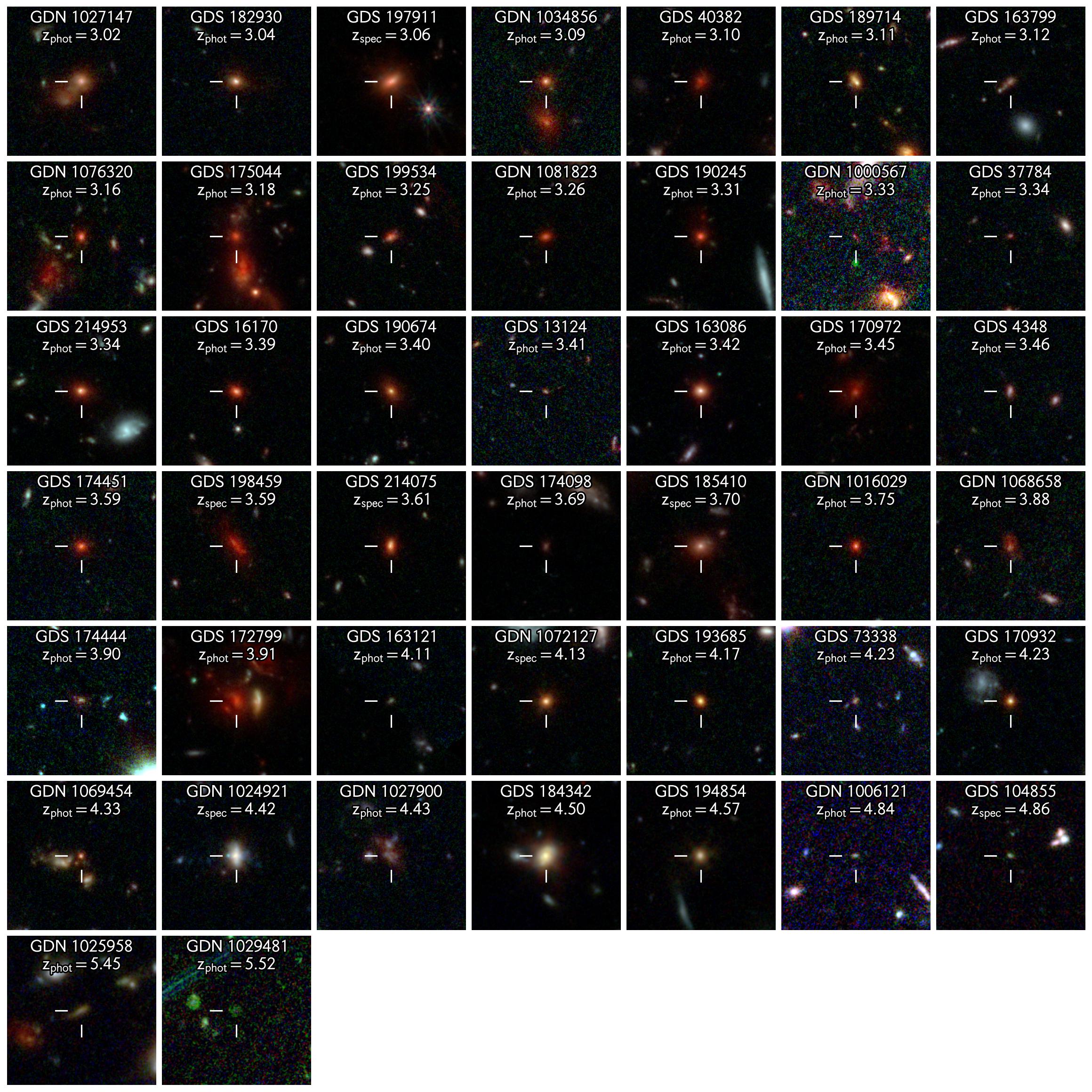}
    \caption{This figure shows false-color rest-frame optical images of the candidates. Each galaxy in the final sample (Section \ref{sec:sample_qg}) is shown in its own panel, which displays a 6\arcsec x 6\arcsec postage stamp and is labeled with its ID and redshift. The blue, green, and red channels of each image correspond to the approximate rest-frame $U$, $B$, and $R$ bandpasses, respectively. The candidates are generally red in the rest-frame optical and present compact morphologies.}
    \label{fig:rgb_images}
\end{figure*}

\subsection{Spectral Energy Distribution Fitting Using {\sc bagpipes}}
\label{sec:bagpipes}

In order to obtain the stellar masses and SFRs needed to identify quiescent galaxies, we fit the SEDs of the galaxies in the observational sample that are pre-selected. In Section \ref{sec:sample_qg}, two ways of pre-selecting are described. We use the Bayesian SED-fitting tool {\sc bagpipes} \citep{Carnall18} to fit the combined HST+JWST SED of each galaxy that is pre-selected. {\sc bagpipes} fits SEDs using the {\sc MultiNest} Markov Chain Monte Carlo algorithm \citep{Feroz09} and returns marginalized posterior probability distribution functions for each parameter of interest. 

We adopt the priors listed in Table 1 in \citet{Carnall2020}, although we adjust those on the metallicity and dust attenuation in the rest-frame $V$-band, $A_V$. This is done based on arguments for physically motivated priors given in Sec. 6 by \citet{delavega25}. We assume a Gaussian prior on the metallicity $Z/Z_{\odot}$ with a mean $\mu=0.3$ and standard deviation $\sigma=0.5$. We assume another Gaussian prior on the rest-frame $V$-band attenuation, $A_V$, with $\mu=0.3$, $\sigma=1.0$, and range of [0, 6]. This prior is adopted from \citet{Leja19}. 

Other priors are as follows. A uniform prior on the log of the stellar mass $\log\left(M_{\star}/M_{\odot}\right)$ is assumed over the range of 1 to 13. A flexible attenuation curve is assumed and follows the model developed by \citet{Salim18}. This model modifies the shape of the attenuation curve measured by \citet{Calzetti00} by multiplying it by a power law that varies with wavelength. A prior on the slope of the power law, $\delta$, is set to be Gaussian with $\mu=0$ and $\sigma=0.1$ over the range [-0.3, 0.3]. The \citet{Calzetti00} curve is obtained by setting $\delta = 0$. The strength of the dust absorption feature at rest-frame 2,175 \AA \ is also modeled by \citet{Salim18}. We set a uniform prior on it over the range [0, 5]. Another prior is assumed for the differential attenuation between young ($\leq10$ Myr) and old stars \citep[see, e.g.,][]{Calzetti00, CF00}. The former are generally dustier than the latter due to additional attenuation from birth clouds. A uniform prior on the ratio of $A_V$ for young stars and that for old stars is assumed over the range of 1 to 5. Nebular emission is added to stellar emission during the fits. A uniform prior is set on the log of the ionization parameter $\log U$ over the range [-4, -2]. The nebular metallicity is assumed to be the same as that of the stars. 

A crucial component of our SED fits is the assumption of a double power-law star-formation history (SFH; see also \citealt{Alberts24} and \citealt{Russell24}). It is given by 
\begin{equation}
    \textrm{SFR}(t) \propto \left[ \left(\frac{t}{\tau}\right)^{\alpha} + \left(\frac{t}{\tau}\right)^{-\beta} \right]^{-1},
\end{equation}
where $t$ is the time since star-formation began in a galaxy, $\tau$ is the time when the SFH peaks since the Big Bang, $\alpha$ is the declining slope, and $\beta$ is the rising slope. This SFH has been shown to accurately reproduce SFHs in cosmological hydrodynamical simulations and it allows for SFHs with rapid cutoffs in SFR with time, which is needed to accurately model the SEDs of high-redshift quiescent galaxies \citep{Pacifici16, Carnall18, Park23}. We assume log-uniform priors for $\alpha$ and $\beta$, each of which is over the range [0.01, 1000]. A uniform prior on $\tau$ is set and ranges from 0.1 Gyr to the age of the Universe at the observed redshift of a given galaxy. 

During the fit, the redshift is fixed to the spectroscopic redshift, otherwise, the photometric redshift from JADES is assumed. In the larger candidate pool we create (Section \ref{sec:preselect_radial}), 17\% (387/2,282) of galaxies have spectroscopic redshifts. For a very small subset (19/2,282), photometric redshifts were fit with {\sc bagpipes}. We apply upper limits on photometry with S/N $< 3$ following the procedure by \citet{delavega25}.

Stellar population synthesis templates come from version 2.2.1 of the Binary Population and Spectral Synthesis \citep[BPASS;][]{Eldridge17, StanwayEldridge18} models. Nebular emission is modeled using the {\sc cloudy} \citep{Ferland17} photoionization code. The initial mass function is set to the default `135\_300' model listed in Table 2 in \citet{StanwayEldridge18}.

\section{The Sample of High-Redshift Quiescent Galaxies}
\label{sec:sample_qg}

We now identify high-redshift quiescent galaxies by pre-selecting candidate pools in the UMAP parameter space. Two different ways to perform the pre-selection are considered (Sections \ref{sec:preselect_radial} and \ref{sec:preselect_rectangle}). One way is more comprehensive than the other and yields a final sample of 44 quiescent galaxy candidates. This sample is listed in Table \ref{tab:qgs}. In Section \ref{sec:overlap}, we assess the overlap of our final sample with those from previous studies. There are 17 candidates in common with the literature and 27 are newly discovered. We comment on the efficiency of our technique compared with that of another observed-frame color-based technique in Section \ref{sec:efficiency}. We calculate number densities in Section \ref{sec:number_densities}. Number densities of quiescent galaxies  identified in this work over $3<z<6$ are similar to those found by other studies.

We note that the exact method used to pre-select galaxies in UMAP space is arbitrary. Other ways of pre-selecting are possible, which may be more efficient that the two presented here. 

\begin{figure*}
    \includegraphics[width=\textwidth]{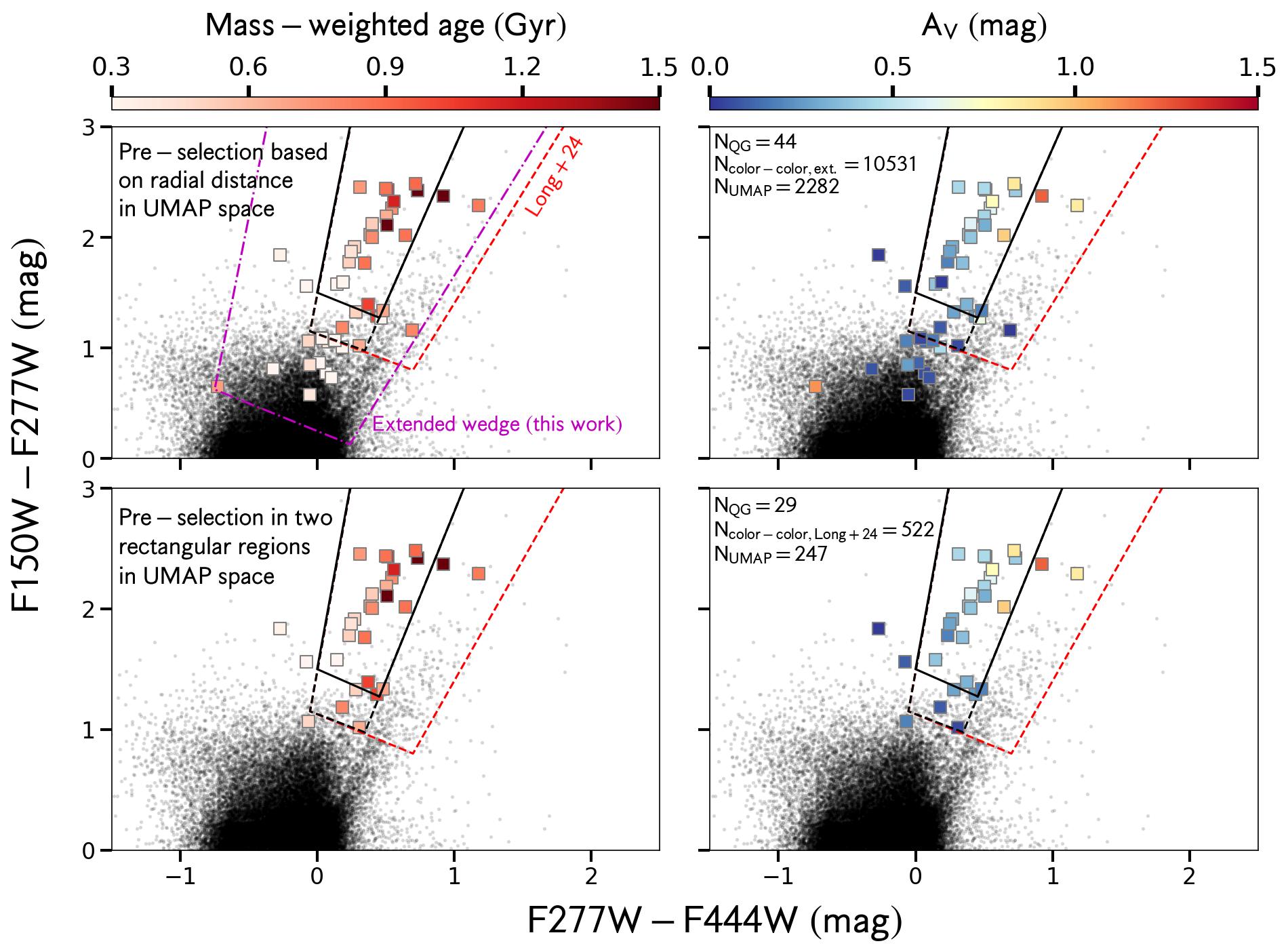}
    \caption{We find that it is more efficient to pre-select a sample of high-redshift galaxies using the UMAP technique than it is using a single color-color diagram. This figure shows the results of two pre-selection techniques in the F150W--F277W vs. F277W--F444W color-color diagram \citep{Long24}. Galaxies in the observational sample are shown as small dots. Candidate quiescent galaxies are shown as squares colored by mass-weighted age on the left and $A_V$ on the right. Redder candidates are generally older and dustier, regardless of the pre-selection technique. In the top row, we show the final sample of 44 candidates. They were pre-selected based on the radial distance to a quiescent model in UMAP parameter space. A total of 2,282 galaxies was part of this candidate pool. If one were to select all 44 candidates using color-color cuts, one would end up with the extended wedge, shown in magenta dash-dot lines, and end up pre-selecting a sample of 10,531 galaxies. This is nearly five times larger than the pre-selection using UMAP. In the bottom row, we show 29 candidates that were pre-selected using two rectangular regions in UMAP space. This technique pre-selected 247 galaxies in total. Nearly all candidates reside within the color-color wedges proposed by \citet{Long24}, which are shown as solid, black dashed, and red dashed lines. Pre-selecting galaxies using these wedges increases the pool to 522, which is more than double that found using UMAP. Most of the candidates that are not selected using the \citet{Long24} color-color criteria are young and have mass-weighted ages of $\sim300$ Myr. }
    \label{fig:long24_diagram}
\end{figure*}

\begin{figure*}
    \includegraphics[width=\textwidth]{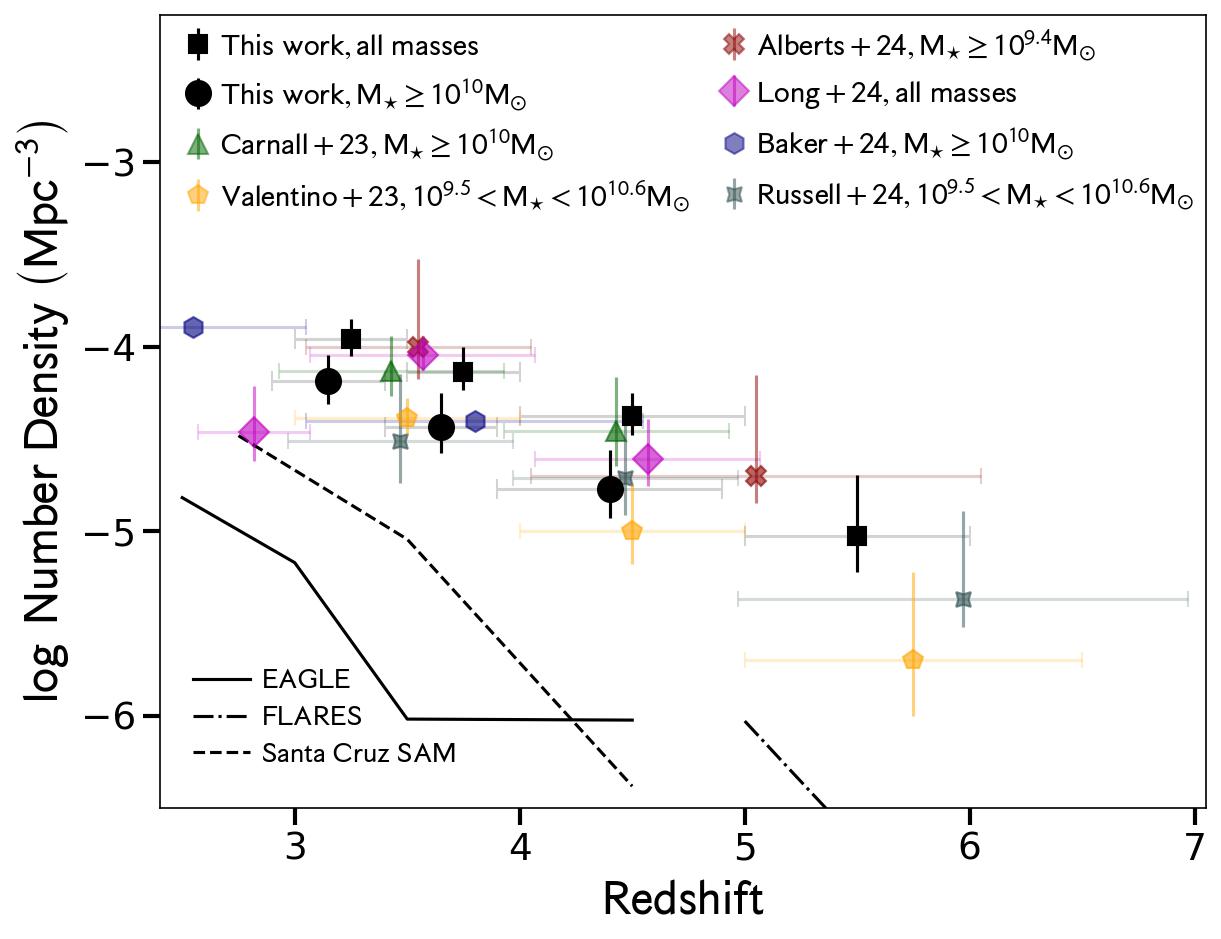}
    \caption{This figure shows comoving number densities of quiescent galaxies at $3<z<6$ using the UMAP technique (Table \ref{tab:number_densities}). Measurements for the final sample of quiescent galaxies identified in this work (Table \ref{tab:qgs}) are done in two mass ranges: at all masses (black squares) and at $M_{\star} \geq 10^{10}M_{\odot}$ (black dots). These are compared with measurements from the literature based on photometric samples \citep{Carnall23_MNRAS, Valentino23, Alberts24, Long24, Russell24} and those based on spectroscopy \citep{Baker24}. Previous measurements are shown as various symbols and are offset slightly in redshift for clarity. Number densities in this work agree well with earlier studies at $z<4$ and are generally higher at $z>4$, although still consistent within uncertainties.
    As was found in earlier studies, observed number densities exceed those from simulations by $\sim1-2$ dex. In this figure are shown measurements from the EAGLE \citep[][solid line]{Crain15, Schaye15, Valentino20} and FLARES \citep[][dash-dot line]{Lovell21, Vijayan21} simulations and the Santa Cruz semi-analytic model \citep[][dashed line]{Somerville15, Somerville21, Yung19, Yung22}. }
    \label{fig:number_density}
\end{figure*}

\subsection{Pre-Selection Based on Radial Distance From Quiescent Models}
\label{sec:preselect_radial}

The first way we pre-select galaxies is by choosing all galaxies that lie within a radial distance of 1 in UMAP parameter space from any high-redshift quiescent galaxy model (Section \ref{sec:def_qg}). A value of 1 is chosen because the spectroscopically confirmed quiescent galaxies in the observational sample (Figure \ref{fig:umap_jades_baker} and Section \ref{sec:baker24}) lie within a radius of 0.5 and this is doubled to ensure we capture as many quiescent galaxies as possible. This method is more comprehensive than that described in the next section. 

In Figure \ref{fig:umap_radial_selection}, the candidate pool of pre-selected galaxies is shown to the left. This pool consists of 2,282 galaxies, which are shown as dots colored by their F150W-F277W color. Of these, we find 44 quiescent galaxy candidates, which are plotted as red squares. They satisfy both the sSFR criterion in Section \ref{sec:def_qg} and the following redshift criteria. If the redshift is spectroscopic, it must lie between $z=3$ and $z=6$. If the redshift is photometric, the median of its redshift probability distribution function (PDF) must lie between $z=3$ and $z=6$. Further, the 16th percentile of the PDF must lie at $z\geq2.5$ and its 84th percentile must lie at $z\leq6.5$. 

The quiescent galaxy candidates from this pre-selection method are listed in Table \ref{tab:qgs}. Their coordinates, stellar masses, sSFRs, mass-weighted ages, and dust attenuation values are also listed. 

\subsection{Pre-Selection Based on Rectangular Cuts}
\label{sec:preselect_rectangle}

Most of the quiescent galaxy candidates are in proximity to two groups of high-redshift quiescent galaxy models (Regions A and C). These groups are mentioned earlier in Section \ref{sec:2d_viz} and include all but two of the high-redshift quiescent models. These regions are plotted in the right panel of Figure \ref{fig:umap_radial_selection}. The rectangular region that surrounds the bulk of the models is given by $-3.5 \leq \textrm{UMAP 1} \leq -3.1, -0.35 \leq \textrm{UMAP 2} \leq -0.1$. The other, which surrounds only a handful of models, is given by $-4.0 \leq \textrm{UMAP 1} \leq -2.8, -4.8 \leq \textrm{UMAP 2} \leq -4.6$. 

There are 247 galaxies in these rectangular cuts. These include 29 candidates, nearly two-thirds of the final sample. 

\subsection{Overlap with Previous Studies}
\label{sec:overlap}

We match the final sample of quiescent galaxies with catalogs of quiescent galaxy candidates at $z\geq3$ from \citet{Merlin19}, \citet{Carnall2020}, \citet{Shahidi20}, \citet{Alberts24}, \citet{Baker24}, and \citet{Russell24}. The first three are based on pre-JWST observations, the fourth and sixth are based on both JWST/NIRCam and JWST/MIRI photometry, and the fifth is based on JWST/NIRSpec spectroscopy. A search radius of 0.5\arcsec \ is used. There are 17 galaxies that have been previously identified as quiescent candidates in the literature (see rightmost column in Table \ref{tab:qgs}). All of these matches have stellar masses greater than $10^{10}~M_{\odot}$. This is because three of these studies sought out only massive quiescent galaxies. 

\subsection{Properties of the Candidates}
We now describe the candidate quiescent galaxies in terms of their stellar masses, sSFRs, mass-weighted ages, and dust content, and compare with other studies. We then briefly discuss their morphologies based on visual inspection of false-color images shown in Figure \ref{fig:rgb_images}. 

The candidates range in stellar mass from $10^{8.52}$ to $10^{11.63}~M_{\odot}$ and the median value is $10^{10.3}~M_{\odot}$. The sSFRs span a wide range, $-67.93 < \log \left(\textrm{sSFR/yr}^{-1}\right)<-9.88$, and the median is $-14.66$. We note that the lowest sSFR values are difficult to constrain and are accompanied with large uncertainties \citep[see, e.g., Sec. 6.2 in][]{Pacifici23}. Mass-weighted ages span 0.13 Gyr to 1.65 Gyr, with the median equal to 0.58 Gyr. Finally, the dust content, given by the attenuation in the rest-frame $V$-band, $A_V$, ranges from 0.01 to 1.23 mag and its median is 0.29 mag. Age and dust are both correlated with stellar mass such that more massive galaxies tend to be older and dustier. 

The properties of the quiescent candidates presented here are similar to those measured in other studies. Similar ranges in stellar mass are found by \citet{Valentino23}, \citet{Alberts24}, \citet{Long24}, and \citet{Russell24}. The candidates in these studies have stellar masses as low as $10^{8.5} - 10^{8.8}~M_{\odot}$, in agreement with the lowest masses found in this work. A wide range in log sSFR is also found by \citet{Alberts24} and \citet{Russell24}, both of which use double power-law SFHs. \citet{Alberts24} reported mass-weighted ages in the range of $\sim0.2-0.8$ Gyr, which overlaps well with the range in age measured in this work. Ranges in $A_V$ are not found in most studies. However, quiescent galaxies over $3\lesssim z<6$ generally have little dust ($A_V < 0.5$ mag) and occasional individual galaxies $A_V \gtrsim 1$ mag are found \citep{Alberts24, Nanayakkara24, Setton24, Siegel24}. This is in qualitative agreement with what is reported here. 

The candidates are generally red in the rest-frame optical and exhibit compact morphologies. In Figure \ref{fig:rgb_images}, false-color rest-frame optical images of the candidates are shown. Each candidate is shown in its own panel. While most galaxies appear red and compact, some display disk-like or edge-on morphologies (e.g., 1025958) and others may be interacting with neighboring galaxies (e.g., 1027147, 172799, and 1069454). Further analysis of the morphologies of quiescent galaxies over $3<z<6$ will be conducted in future work. 

\subsection{Efficiency in Selecting High-Redshift Galaxies Using the Technique Developed in This Work}
\label{sec:efficiency}

To assess the efficiency in pre-selecting candidates using UMAP, we compare with what is found using the color-color diagram developed by \citet{Long24}, F150W--F277W vs. F277W--F444W.  That work also proposed various color cuts. We compare the efficiency for each of the two pre-selection techniques above with an equivalent cut in color-color space. These comparisons are shown in Figure \ref{fig:long24_diagram}.

For the pre-selection based on radial distance, a total of 2,282 galaxies was in the candidate pool, which includes the 44 candidates in the final sample. If one were to select all of the candidates using color-color cuts, one would end up with a color-color wedge that extends far past the cuts proposed by \citet{Long24}. This wedge is shown as magenta dash-dot lines in Figure \ref{fig:long24_diagram}. It was found by adjusting only the intercepts in the color-color criteria provided by \citet{Long24}. This wedge includes 10,531 galaxies, which is nearly five times larger than the pre-selection using UMAP. If one were to ignore the outlier candidate at F277W--F444W$\sim-0.5$ mag, the wedge would become more compact and would include $\sim4,500$ galaxies. This is still considerably larger than what is pre-selected using UMAP.  

For the pre-selection based on the rectangular cuts, a total of 247 galaxies was chosen. This includes 29 candidates, nearly all of which lie within the ``wide'' wedge developed by \citet{Long24}. Pre-selecting galaxies using this wedge increases the pool to 522, which is more than double that found using UMAP. 

It is stressed that, while pre-selection using UMAP is more efficient than a single color-color wedge, the former is more complicated to execute than the latter in practice. In the regime of very large sample sizes, pre-selection via UMAP is an attractive option. For smaller samples, such as individual deep JWST surveys, the color-color cuts proposed by \citet{Long24} already perform efficiently. 

A significant advantage of pre-selecting using UMAP is the efficiency with which young ($\lesssim300$ Myr) quiescent galaxies are identified. Quiescent galaxies that are pre-selected using the \citet{Long24} criteria tend to have mass-weighted ages of $\gtrsim0.5$ Gyr. Nearly all of the 14 candidates in this work that lie beyond the \citet{Long24} criteria have younger ages. However, pre-selecting these candidates using color-color cuts would entail fitting the SEDs of much larger samples (by a factor of $\sim5$) than what would be pre-selected using UMAP. 

\begin{deluxetable}{c c c c}[t!]
\tablecaption{Number Densities of Quiescent Galaxies Identified in This Work \label{tab:number_densities}}
\tablewidth{\textwidth}
\tablehead{Redshift range & $\log\left(M_{\star}\right)$ & Raw number & Number density \\
     & $\left(M_{\odot}\right)$ & & $\left(\textrm{Mpc}^{-3}\right)$ \\}
     \startdata
     $3.0\leq z < 3.5$ & all & 21 & $1.1^{+0.3}_{-0.3}\times10^{-4}$ \\
     & $\geq 10$ & 14 & $6.5^{+2.5}_{-2.1}\times 10^{-5}$ \\
     $3.5 \leq z < 4.0$ & all & 9 & $7.3^{+2.7}_{-1.9}\times10^{-5}$ \\
     & $\geq 10$ & 6 & $3.7^{+2.0}_{-1.4}\times 10^{-5}$ \\
     $4.0\leq z < 5.0$ & all & 12 & $3.9^{+1.4}_{-1.0}\times 10^{-5}$ \\
     & $\geq 10$ & 6 & $1.7^{+1.1}_{-0.7}\times 10^{-5}$ \\
     $5.0\leq z \leq 6.0$ & all & 2 & $0.9^{+1.1}_{-0.5}\times 10^{-5}$ \\
     & $\geq 10$ & 0 & $0.0^{+0.7}_{-0.0}\times 10^{-5}$ \\
     \hline
     \enddata
\end{deluxetable}

\subsection{Number Densities}
\label{sec:number_densities}

Number densities for the final sample are derived over the redshift ranges $3\leq z<3.5, 3.5\leq z<4, 4\leq z<5,$ and $5\leq z\leq6$. Following \citet{Long24}, uncertainties on the number densities are obtained by running 1,000 Monte Carlo realizations. In each realization, the redshift of a galaxy is randomly perturbed assuming a uniform distribution with a minimum set to its lower bound on uncertainty and a maximum set to its upper bound. In the case of photometric redshifts, these are assumed to be the 16th and 84th percentiles of the redshift PDF. For spectroscopic redshifts, uncertainties are assumed to be $\pm0.005$. The 16th and 84th percentiles from the 1,000 Monte Carlo iterations are taken to be the lower and upper bounds on the number density uncertainties. We add Poisson noise \citep{Gehrels86} in quadrature. Number densities and their uncertainties are reported in Table \ref{tab:number_densities} for all candidates and for those with stellar mass $M_{\star}\geq10^{10}~M_{\odot}$. 

The number densities for the final sample are shown in Figure \ref{fig:number_density}. They are plotted in two mass regimes: at all masses (black squares) and at $M_{\star}\geq10^{10}M_{\odot}$ (black dots). The former are higher than the latter by about a factor of two at all redshifts, except at the highest redshifts, where no massive candidates are found. Both sets of measurements show the same trend with redshift: number densities are lowest at the highest redshifts (number density $n\sim10^{-5}~\textrm{Mpc}^{-3}$ at $z\sim5.5$) and increase with decreasing redshift ($n\sim10^{-4}$ at $z\sim3.25$).

We now compare measurements in this work with previous ones from studies based on observations and simulations. Observed number densities from Table 3 in \citet{Carnall23_MNRAS}, Table 2 in \citet{Valentino23}, Sec. 5.4 in \citet{Alberts24}, Table 2 in \cite{Long24}, Table 1 in \citet{Baker24}, and Table 1 in \citet{Russell24} are shown in Figure \ref{fig:number_density} as various symbols. The first four studies are based on photometric samples and the last on spectroscopy. For the comparison with \citet{Carnall23_MNRAS}, \citet{Alberts24}, and \citet{Russell24}, we use the ``robust'' sub-sample in each work. For that with \citet{Valentino23}, we use their measurements based on the ``$UVJ$ padded'' selection technique, which gives the highest number densities. There is good agreement with these works at $z<4$. Number densities measured in this work are in general higher than those from earlier studies at $z>4$, although there is consistency within uncertainties. We compare with simulated number densities from the EAGLE \citep[][]{Crain15, Schaye15, Valentino20} and FLARES \citep[][]{Lovell21, Vijayan21} simulations and the Santa Cruz semi-analytic model \citep[][]{Somerville15, Somerville21, Yung19, Yung22}. These are shown as lines in various styles in the figure. Our number densities are about 1-2 dex higher, as was found by other studies based on observations. 

There is nascent evidence for environment-dependent quenching at early times, which may affect the interpretation of the number densities measured here and in other studies. Further spectroscopic observations are necessary to confirm this. The works by \citet{Carnall23_MNRAS}, \citet{Valentino23}, \cite{Long24}, and \citet{Russell24} include galaxies in the Cosmic Evolution Early Release Science Survey \citep[CEERS,][]{Finkelstein22, Finkelstein23, Finkelstein25, Bagley23} field, which contains an overdensity at $z\approx3.4$ that was recently confirmed with JWST spectroscopy (\citealt{Jin24}, see also Sec. 4.3 in \citealt{Valentino23}). Two galaxies in the overdensity are massive quiescent galaxies and there are $\sim200$ candidate members with photometric redshifts over $3.3<z<3.6$. In the GOODS-S field, \citet{Alberts24} found two robust quiescent candidates in an overdensity at $z\sim3.4$ and three tentative candidates in another overdensity at $z\sim3.7$ (see their Sec. 5.3 and Appendix B). Both overdensities overlap on the sky with spectroscopically confirmed massive protostructures discovered by \citet{Shah24} at similar redshifts. 

\section{Conclusions}
\label{sec:conclusions}

We use JWST and HST photometry from the JADES survey in the GOODS-N and GOODS-S fields to search for quiescent galaxies over $3<z<6$. A pre-selection is performed using the UMAP machine-learning technique such that only a fraction of our observational sample of 43,926 galaxies needs to be combed for quiescent candidates. A final sample of 44 candidates is obtained (Table \ref{tab:qgs} and Figure \ref{fig:rgb_images}), of which 27 are new discoveries. 

The main conclusions from this work are summarized as follows:
\begin{enumerate}
    \item We train the UMAP machine-learning algorithm on seven NIRCam observed-frame colors, which were chosen to separate high-redshift quiescent galaxies from other galaxies (Section \ref{sec:training_colors}). Colors from $\approx75,000$ JAGUAR model galaxies \citep{Williams18} are used to produce a two-dimensional visualization via UMAP. Model $z\geq3$ quiescent galaxies are tightly clustered in the UMAP parameter space.
    \item An observational sample of 43,926 galaxies is selected in JADES with magnitudes brighter than 29th in several bandpasses (Section \ref{sec:data}). The sample is mapped onto the UMAP visualization (Section \ref{sec:identify_qgs}). There are 17 galaxies in the sample that are spectroscopically confirmed quiescent galaxies at $z\gtrsim3$ \citep{Baker24}. These are grouped closely together with the model quiescent galaxies. 
    \item Pre-selection of quiescent galaxies is performed in UMAP parameter space in two ways (Section \ref{sec:sample_qg}): (1) choosing galaxies within a certain radial distance from quiescent models, and (2) choosing galaxies within two rectangular regions that enclose the bulk of the $z\geq3$ quiescent models. For the former method, the combined HST and JWST spectral energy distributions, comprising as many as 17 bandpasses, are fit for 2,282 galaxies. This results in 44 quiescent candidates, which make up the final sample in this work. For the latter, 247 galaxies are examined and 29 candidates are found, all of which are in the final sample. There are 17 previously identified quiescent candidates in the final sample, leaving 27 new discoveries.
    \item Pre-selection via our technique is more efficient than traditional color-color pre-selection methods. Most candidates found through the first pre-selection technique can be found using the observed color-color cuts proposed by \citet{Long24}. However, there are 14 that are not selected according to these criteria. These are generally young quiescent galaxies with mass-weighted ages of $\lesssim300$ Myr. If more inclusive color-color criteria were used to select all 44 candidates, about 10,500 galaxies would be chosen, in contrast to the nearly 2,300 chosen using UMAP. Of the 29 candidates pre-selected in the two rectangular regions above, all but three lie in the color-color cuts proposed by \citet{Long24}. If pre-selection using these cuts is applied, 522 galaxies are chosen. If pre-selected using UMAP, 247 galaxies are chosen. 
    \item Number densities of our quiescent candidates at $z<4$ are consistent with studies based on observations. At $z>4$, higher number densities are found here than in previous works in general, but there is consistency within uncertainties. As was found in earlier studies, our observed number densities are $\sim1-2$ dex higher than those measured in simulations. 
\end{enumerate}

This study builds upon previous efforts to identify elusive high-redshift quiescent galaxies using the unparalleled combination of deep HST and JWST photometry in JADES. Future work may benefit from applying the technique developed here to other JWST datasets, such as CEERS \citep{Finkelstein22, Finkelstein23, Bagley23}, NEXUS \citep{Shen24}, PANORAMIC \citep{Williams24}, PRIMER \citep{Dunlop21}, and UNCOVER \citep{Bezanson24}. We provide {\tt python} code to train the UMAP algorithm and reproduce Figures \ref{fig:umap_ssfr_z_tauv} through \ref{fig:umap_radial_selection} and the catalog of {\sc bagpipes} fits (Section \ref{sec:bagpipes}) at \href{https://github.com/alex-delavega/umap_jades_2025}{https://github.com/alex-delavega/umap\_jades\_2025}. 

\section*{Acknowledgements}
We dedicate this paper to Mitchell D. Babcock, who passed away during the drafting of this manuscript. He was an outstanding student and friend. His intense curiosity in the early stages of this project helped reveal the feasibility of our technique and inspired us to continue his work after his passing. He left us far too early. We will miss him dearly. 

M.D.B. and D.A.R. were supported by U.S. National Science Foundation (NSF) Research Experience for Undergraduates (REU) grant AST-2244610.

We thank the JADES team for designing and preparing their observations and releasing a rich public dataset. We also thank the STScI staff for enabling this science.

All of the JWST data used in this paper can be found in {\it MAST}: (\dataset[10.17909/8tdj-8n28]{http:/dx.doi.org/10.17909/8tdj-8n28} and \dataset[10.17909/fsc4-dt61]{http:/dx.doi.org/10.17909/fsc4-dt61}), as can all of the HST data used in this paper: \dataset[10.17909/T94S3X]{http:/dx.doi.org/10.17909/T94S3X}. 

\software{astropy \citep{2013A&A...558A..33A,2018AJ....156..123A,Astropy22}, 
matplotlib \citep{Hunter:2007},
numpy \citep{harris2020array},
UMAP \citep{McInnes18}, 
scikit-learn \citep{scikit-learn},
scipy \citep{2020SciPy-NMeth}
          }






\bibliography{sample631}{}

\begin{thebibliography}{}
\expandafter\ifx\csname natexlab\endcsname\relax\def\natexlab#1{#1}\fi
\providecommand{\url}[1]{\href{#1}{#1}}
\providecommand{\dodoi}[1]{doi:~\href{http://doi.org/#1}{\nolinkurl{#1}}}
\providecommand{\doeprint}[1]{\href{http://ascl.net/#1}{\nolinkurl{http://ascl.net/#1}}}
\providecommand{\doarXiv}[1]{\href{https://arxiv.org/abs/#1}{\nolinkurl{https://arxiv.org/abs/#1}}}

\bibitem[{{Ackermann} {et~al.}(2018){Ackermann}, {Schawinski}, {Zhang},
  {Weigel}, \& {Turp}}]{Ackermann18}
{Ackermann}, S., {Schawinski}, K., {Zhang}, C., {Weigel}, A.~K., \& {Turp},
  M.~D. 2018, \mnras, 479, 415, \dodoi{10.1093/mnras/sty1398}

\bibitem[{{Alberts} {et~al.}(2024){Alberts}, {Williams}, {Helton}, {Suess},
  {Ji}, {Shivaei}, {Lyu}, {Rieke}, {Baker}, {Bonaventura}, {Bunker},
  {Carniani}, {Charlot}, {Curtis-Lake}, {D'Eugenio}, {Eisenstein}, {de Graaff},
  {Hainline}, {Hausen}, {Johnson}, {Maiolino}, {Parlanti}, {Rieke},
  {Robertson}, {Sun}, {Tacchella}, {Willmer}, \& {Willott}}]{Alberts24}
{Alberts}, S., {Williams}, C.~C., {Helton}, J.~M., {et~al.} 2024, \apj, 975,
  85, \dodoi{10.3847/1538-4357/ad66cc}

\bibitem[{{Antwi-Danso} {et~al.}(2023){Antwi-Danso}, {Papovich}, {Leja},
  {Marchesini}, {Marsan}, {Martis}, {Labb{\'e}}, {Muzzin}, {Glazebrook},
  {Straatman}, \& {Tran}}]{AntwiDanso23_ugi}
{Antwi-Danso}, J., {Papovich}, C., {Leja}, J., {et~al.} 2023, \apj, 943, 166,
  \dodoi{10.3847/1538-4357/aca294}

\bibitem[{{Astropy Collaboration} {et~al.}(2013){Astropy Collaboration},
  {Robitaille}, {Tollerud}, {Greenfield}, {Droettboom}, {Bray}, {Aldcroft},
  {Davis}, {Ginsburg}, {Price-Whelan}, {Kerzendorf}, {Conley}, {Crighton},
  {Barbary}, {Muna}, {Ferguson}, {Grollier}, {Parikh}, {Nair}, {Unther},
  {Deil}, {Woillez}, {Conseil}, {Kramer}, {Turner}, {Singer}, {Fox}, {Weaver},
  {Zabalza}, {Edwards}, {Azalee Bostroem}, {Burke}, {Casey}, {Crawford},
  {Dencheva}, {Ely}, {Jenness}, {Labrie}, {Lim}, {Pierfederici}, {Pontzen},
  {Ptak}, {Refsdal}, {Servillat}, \& {Streicher}}]{2013A&A...558A..33A}
{Astropy Collaboration}, {Robitaille}, T.~P., {Tollerud}, E.~J., {et~al.} 2013,
  \aap, 558, A33, \dodoi{10.1051/0004-6361/201322068}

\bibitem[{{Astropy Collaboration} {et~al.}(2018){Astropy Collaboration},
  {Price-Whelan}, {Sip{\H{o}}cz}, {G{\"u}nther}, {Lim}, {Crawford}, {Conseil},
  {Shupe}, {Craig}, {Dencheva}, {Ginsburg}, {VanderPlas}, {Bradley},
  {P{\'e}rez-Su{\'a}rez}, {de Val-Borro}, {Aldcroft}, {Cruz}, {Robitaille},
  {Tollerud}, {Ardelean}, {Babej}, {Bach}, {Bachetti}, {Bakanov}, {Bamford},
  {Barentsen}, {Barmby}, {Baumbach}, {Berry}, {Biscani}, {Boquien}, {Bostroem},
  {Bouma}, {Brammer}, {Bray}, {Breytenbach}, {Buddelmeijer}, {Burke},
  {Calderone}, {Cano Rodr{\'\i}guez}, {Cara}, {Cardoso}, {Cheedella}, {Copin},
  {Corrales}, {Crichton}, {D'Avella}, {Deil}, {Depagne}, {Dietrich}, {Donath},
  {Droettboom}, {Earl}, {Erben}, {Fabbro}, {Ferreira}, {Finethy}, {Fox},
  {Garrison}, {Gibbons}, {Goldstein}, {Gommers}, {Greco}, {Greenfield},
  {Groener}, {Grollier}, {Hagen}, {Hirst}, {Homeier}, {Horton}, {Hosseinzadeh},
  {Hu}, {Hunkeler}, {Ivezi{\'c}}, {Jain}, {Jenness}, {Kanarek}, {Kendrew},
  {Kern}, {Kerzendorf}, {Khvalko}, {King}, {Kirkby}, {Kulkarni}, {Kumar},
  {Lee}, {Lenz}, {Littlefair}, {Ma}, {Macleod}, {Mastropietro}, {McCully},
  {Montagnac}, {Morris}, {Mueller}, {Mumford}, {Muna}, {Murphy}, {Nelson},
  {Nguyen}, {Ninan}, {N{\"o}the}, {Ogaz}, {Oh}, {Parejko}, {Parley}, {Pascual},
  {Patil}, {Patil}, {Plunkett}, {Prochaska}, {Rastogi}, {Reddy Janga},
  {Sabater}, {Sakurikar}, {Seifert}, {Sherbert}, {Sherwood-Taylor}, {Shih},
  {Sick}, {Silbiger}, {Singanamalla}, {Singer}, {Sladen}, {Sooley},
  {Sornarajah}, {Streicher}, {Teuben}, {Thomas}, {Tremblay}, {Turner},
  {Terr{\'o}n}, {van Kerkwijk}, {de la Vega}, {Watkins}, {Weaver}, {Whitmore},
  {Woillez}, {Zabalza}, \& {Astropy Contributors}}]{2018AJ....156..123A}
{Astropy Collaboration}, {Price-Whelan}, A.~M., {Sip{\H{o}}cz}, B.~M., {et~al.}
  2018, \aj, 156, 123, \dodoi{10.3847/1538-3881/aabc4f}

\bibitem[{{Astropy Collaboration} {et~al.}(2022){Astropy Collaboration},
  {Price-Whelan}, {Lim}, {Earl}, {Starkman}, {Bradley}, {Shupe}, {Patil},
  {Corrales}, {Brasseur}, {N{\"o}the}, {Donath}, {Tollerud}, {Morris},
  {Ginsburg}, {Vaher}, {Weaver}, {Tocknell}, {Jamieson}, {van Kerkwijk},
  {Robitaille}, {Merry}, {Bachetti}, {G{\"u}nther}, {Aldcroft},
  {Alvarado-Montes}, {Archibald}, {B{\'o}di}, {Bapat}, {Barentsen},
  {Baz{\'a}n}, {Biswas}, {Boquien}, {Burke}, {Cara}, {Cara}, {Conroy},
  {Conseil}, {Craig}, {Cross}, {Cruz}, {D'Eugenio}, {Dencheva}, {Devillepoix},
  {Dietrich}, {Eigenbrot}, {Erben}, {Ferreira}, {Foreman-Mackey}, {Fox},
  {Freij}, {Garg}, {Geda}, {Glattly}, {Gondhalekar}, {Gordon}, {Grant},
  {Greenfield}, {Groener}, {Guest}, {Gurovich}, {Handberg}, {Hart},
  {Hatfield-Dodds}, {Homeier}, {Hosseinzadeh}, {Jenness}, {Jones}, {Joseph},
  {Kalmbach}, {Karamehmetoglu}, {Ka{\l}uszy{\'n}ski}, {Kelley}, {Kern},
  {Kerzendorf}, {Koch}, {Kulumani}, {Lee}, {Ly}, {Ma}, {MacBride}, {Maljaars},
  {Muna}, {Murphy}, {Norman}, {O'Steen}, {Oman}, {Pacifici}, {Pascual},
  {Pascual-Granado}, {Patil}, {Perren}, {Pickering}, {Rastogi}, {Roulston},
  {Ryan}, {Rykoff}, {Sabater}, {Sakurikar}, {Salgado}, {Sanghi}, {Saunders},
  {Savchenko}, {Schwardt}, {Seifert-Eckert}, {Shih}, {Jain}, {Shukla}, {Sick},
  {Simpson}, {Singanamalla}, {Singer}, {Singhal}, {Sinha}, {Sip{\H{o}}cz},
  {Spitler}, {Stansby}, {Streicher}, {{\v{S}}umak}, {Swinbank}, {Taranu},
  {Tewary}, {Tremblay}, {de Val-Borro}, {Van Kooten}, {Vasovi{\'c}}, {Verma},
  {de Miranda Cardoso}, {Williams}, {Wilson}, {Winkel}, {Wood-Vasey}, {Xue},
  {Yoachim}, {Zhang}, {Zonca}, \& {Astropy Project Contributors}}]{Astropy22}
{Astropy Collaboration}, {Price-Whelan}, A.~M., {Lim}, P.~L., {et~al.} 2022,
  \apj, 935, 167, \dodoi{10.3847/1538-4357/ac7c74}

\bibitem[{{Bagley} {et~al.}(2023){Bagley}, {Finkelstein}, {Koekemoer},
  {Ferguson}, {Arrabal Haro}, {Dickinson}, {Kartaltepe}, {Papovich},
  {P{\'e}rez-Gonz{\'a}lez}, {Pirzkal}, {Somerville}, {Willmer}, {Yang}, {Yung},
  {Fontana}, {Grazian}, {Grogin}, {Hirschmann}, {Kewley}, {Kirkpatrick},
  {Kocevski}, {Lotz}, {Medrano}, {Morales}, {Pentericci}, {Ravindranath},
  {Trump}, {Wilkins}, {Calabr{\`o}}, {Cooper}, {Costantin}, {de la Vega},
  {Hilbert}, {Hutchison}, {Larson}, {Lucas}, {McGrath}, {Ryan}, {Wang}, \&
  {Wuyts}}]{Bagley23}
{Bagley}, M.~B., {Finkelstein}, S.~L., {Koekemoer}, A.~M., {et~al.} 2023,
  \apjl, 946, L12, \dodoi{10.3847/2041-8213/acbb08}

\bibitem[{{Baker} {et~al.}(2024){Baker}, {Lim}, {D'Eugenio}, {Maiolino}, {Ji},
  {Arribas}, {Bunker}, {Carniani}, {Charlot}, {de Graaff}, {Hainline},
  {Looser}, {Lyu}, {Rinaldi}, {Robertson}, {Schaller}, {Schaye}, {Scholtz},
  {Ubler}, {Williams}, {Willmer}, {Willott}, \& {Zhu}}]{Baker24}
{Baker}, W.~M., {Lim}, S., {D'Eugenio}, F., {et~al.} 2024, arXiv e-prints,
  arXiv:2410.14773, \dodoi{10.48550/arXiv.2410.14773}

\bibitem[{{Ball} \& {Brunner}(2010)}]{BallBrunner10}
{Ball}, N.~M., \& {Brunner}, R.~J. 2010, International Journal of Modern
  Physics D, 19, 1049, \dodoi{10.1142/S0218271810017160}

\bibitem[{{Belli} {et~al.}(2019){Belli}, {Newman}, \& {Ellis}}]{Belli19}
{Belli}, S., {Newman}, A.~B., \& {Ellis}, R.~S. 2019, \apj, 874, 17,
  \dodoi{10.3847/1538-4357/ab07af}

\bibitem[{{Bezanson} {et~al.}(2024){Bezanson}, {Labbe}, {Whitaker}, {Leja},
  {Price}, {Franx}, {Brammer}, {Marchesini}, {Zitrin}, {Wang}, {Weaver},
  {Furtak}, {Atek}, {Coe}, {Cutler}, {Dayal}, {van Dokkum}, {Feldmann},
  {F{\"o}rster Schreiber}, {Fujimoto}, {Geha}, {Glazebrook}, {de Graaff},
  {Greene}, {Juneau}, {Kassin}, {Kriek}, {Khullar}, {Maseda}, {Mowla},
  {Muzzin}, {Nanayakkara}, {Nelson}, {Oesch}, {Pacifici}, {Pan}, {Papovich},
  {Setton}, {Shapley}, {Smit}, {Stefanon}, {Taylor}, \&
  {Williams}}]{Bezanson24}
{Bezanson}, R., {Labbe}, I., {Whitaker}, K.~E., {et~al.} 2024, \apj, 974, 92,
  \dodoi{10.3847/1538-4357/ad66cf}

\bibitem[{{Bradley}(2023)}]{Bradley23}
{Bradley}, L. 2023, {astropy/photutils: 1.8.0}, 1.8.0, Zenodo,  Zenodo,
  \dodoi{10.5281/zenodo.7946442}

\bibitem[{{Brammer} {et~al.}(2008){Brammer}, {van Dokkum}, \&
  {Coppi}}]{Brammer08}
{Brammer}, G.~B., {van Dokkum}, P.~G., \& {Coppi}, P. 2008, \apj, 686, 1503,
  \dodoi{10.1086/591786}

\bibitem[{{Brammer} {et~al.}(2011){Brammer}, {Whitaker}, {van Dokkum},
  {Marchesini}, {Franx}, {Kriek}, {Labb{\'e}}, {Lee}, {Muzzin}, {Quadri},
  {Rudnick}, \& {Williams}}]{Brammer11}
{Brammer}, G.~B., {Whitaker}, K.~E., {van Dokkum}, P.~G., {et~al.} 2011, \apj,
  739, 24, \dodoi{10.1088/0004-637X/739/1/24}

\bibitem[{{Bunker} {et~al.}(2024){Bunker}, {Cameron}, {Curtis-Lake},
  {Jakobsen}, {Carniani}, {Curti}, {Witstok}, {Maiolino}, {D'Eugenio},
  {Looser}, {Willott}, {Bonaventura}, {Hainline}, {{\"U}bler}, {Willmer},
  {Saxena}, {Smit}, {Alberts}, {Arribas}, {Baker}, {Baum}, {Bhatawdekar},
  {Bowler}, {Boyett}, {Charlot}, {Chen}, {Chevallard}, {Circosta}, {DeCoursey},
  {de Graaff}, {Egami}, {Eisenstein}, {Endsley}, {Ferruit}, {Giardino},
  {Hausen}, {Helton}, {Hviding}, {Ji}, {Johnson}, {Jones}, {Kumari}, {Laseter},
  {L{\"u}tzgendorf}, {Maseda}, {Nelson}, {Parlanti}, {Perna}, {Rauscher},
  {Rawle}, {Rix}, {Rieke}, {Robertson}, {Rodr{\'\i}guez Del Pino}, {Sandles},
  {Scholtz}, {Sharpe}, {Skarbinski}, {Stark}, {Sun}, {Tacchella}, {Topping},
  {Villanueva}, {Wallace}, {Williams}, \& {Woodrum}}]{Bunker24}
{Bunker}, A.~J., {Cameron}, A.~J., {Curtis-Lake}, E., {et~al.} 2024, \aap, 690,
  A288, \dodoi{10.1051/0004-6361/202347094}

\bibitem[{{Calzetti} {et~al.}(2000){Calzetti}, {Armus}, {Bohlin}, {Kinney},
  {Koornneef}, \& {Storchi-Bergmann}}]{Calzetti00}
{Calzetti}, D., {Armus}, L., {Bohlin}, R.~C., {et~al.} 2000, \apj, 533, 682,
  \dodoi{10.1086/308692}

\bibitem[{{Carnall} {et~al.}(2018){Carnall}, {McLure}, {Dunlop}, \&
  {Dav{\'e}}}]{Carnall18}
{Carnall}, A.~C., {McLure}, R.~J., {Dunlop}, J.~S., \& {Dav{\'e}}, R. 2018,
  \mnras, 480, 4379, \dodoi{10.1093/mnras/sty2169}

\bibitem[{{Carnall} {et~al.}(2020){Carnall}, {Walker}, {McLure}, {Dunlop},
  {McLeod}, {Cullen}, {Wild}, {Amorin}, {Bolzonella}, {Castellano}, {Cimatti},
  {Cucciati}, {Fontana}, {Gargiulo}, {Garilli}, {Jarvis}, {Pentericci},
  {Pozzetti}, {Zamorani}, {Calabro}, {Hathi}, \& {Koekemoer}}]{Carnall2020}
{Carnall}, A.~C., {Walker}, S., {McLure}, R.~J., {et~al.} 2020, \mnras, 496,
  695, \dodoi{10.1093/mnras/staa1535}

\bibitem[{{Carnall} {et~al.}(2023{\natexlab{a}}){Carnall}, {McLeod}, {McLure},
  {Dunlop}, {Begley}, {Cullen}, {Donnan}, {Hamadouche}, {Jewell}, {Jones},
  {Pollock}, \& {Wild}}]{Carnall23_MNRAS}
{Carnall}, A.~C., {McLeod}, D.~J., {McLure}, R.~J., {et~al.}
  2023{\natexlab{a}}, \mnras, 520, 3974, \dodoi{10.1093/mnras/stad369}

\bibitem[{{Carnall} {et~al.}(2023{\natexlab{b}}){Carnall}, {McLure}, {Dunlop},
  {McLeod}, {Wild}, {Cullen}, {Magee}, {Begley}, {Cimatti}, {Donnan},
  {Hamadouche}, {Jewell}, \& {Walker}}]{Carnall23_Nature}
{Carnall}, A.~C., {McLure}, R.~J., {Dunlop}, J.~S., {et~al.}
  2023{\natexlab{b}}, \nat, 619, 716, \dodoi{10.1038/s41586-023-06158-6}

\bibitem[{{Cecchi} {et~al.}(2019){Cecchi}, {Bolzonella}, {Cimatti}, \&
  {Girelli}}]{Cecchi19}
{Cecchi}, R., {Bolzonella}, M., {Cimatti}, A., \& {Girelli}, G. 2019, \apjl,
  880, L14, \dodoi{10.3847/2041-8213/ab2c80}

\bibitem[{{Charlot} \& {Fall}(2000)}]{CF00}
{Charlot}, S., \& {Fall}, S.~M. 2000, \apj, 539, 718, \dodoi{10.1086/309250}

\bibitem[{{Chartab} {et~al.}(2023){Chartab}, {Mobasher}, {Cooray}, {Hemmati},
  {Sattari}, {Ferguson}, {Sanders}, {Weaver}, {Stern}, {McCracken}, {Masters},
  {Toft}, {Capak}, {Davidzon}, {Dickinson}, {Rhodes}, {Moneti}, {Ilbert},
  {Zalesky}, {McPartland}, {Szapudi}, {Koekemoer}, {Teplitz}, \&
  {Giavalisco}}]{Chartab23}
{Chartab}, N., {Mobasher}, B., {Cooray}, A.~R., {et~al.} 2023, \apj, 942, 91,
  \dodoi{10.3847/1538-4357/acacf5}

\bibitem[{{Chevallard} \& {Charlot}(2016)}]{Chevallard16}
{Chevallard}, J., \& {Charlot}, S. 2016, \mnras, 462, 1415,
  \dodoi{10.1093/mnras/stw1756}

\bibitem[{{Cook} {et~al.}(2024){Cook}, {Bandi}, {Philipsborn}, {Loveday},
  {Bellstedt}, {Driver}, {Robotham}, {Bilicki}, {Kaur}, {Tempel}, {Baldry},
  {Gruen}, {Longhetti}, {Iovino}, {Holwerda}, \& {Demarco}}]{Cook24}
{Cook}, T.~L., {Bandi}, B., {Philipsborn}, S., {et~al.} 2024, \mnras,
  \dodoi{10.1093/mnras/stae2389}

\bibitem[{{Crain} {et~al.}(2015){Crain}, {Schaye}, {Bower}, {Furlong},
  {Schaller}, {Theuns}, {Dalla Vecchia}, {Frenk}, {McCarthy}, {Helly},
  {Jenkins}, {Rosas-Guevara}, {White}, \& {Trayford}}]{Crain15}
{Crain}, R.~A., {Schaye}, J., {Bower}, R.~G., {et~al.} 2015, \mnras, 450, 1937,
  \dodoi{10.1093/mnras/stv725}

\bibitem[{{Dahlen} {et~al.}(2013){Dahlen}, {Mobasher}, {Faber}, {Ferguson},
  {Barro}, {Finkelstein}, {Finlator}, {Fontana}, {Gruetzbauch}, {Johnson},
  {Pforr}, {Salvato}, {Wiklind}, {Wuyts}, {Acquaviva}, {Dickinson}, {Guo},
  {Huang}, {Huang}, {Newman}, {Bell}, {Conselice}, {Galametz}, {Gawiser},
  {Giavalisco}, {Grogin}, {Hathi}, {Kocevski}, {Koekemoer}, {Koo}, {Lee},
  {McGrath}, {Papovich}, {Peth}, {Ryan}, {Somerville}, {Weiner}, \&
  {Wilson}}]{Dahlen13}
{Dahlen}, T., {Mobasher}, B., {Faber}, S.~M., {et~al.} 2013, \apj, 775, 93,
  \dodoi{10.1088/0004-637X/775/2/93}

\bibitem[{{Davidzon} {et~al.}(2019){Davidzon}, {Laigle}, {Capak}, {Ilbert},
  {Masters}, {Hemmati}, {Apostolakos}, {Coupon}, {de la Torre}, {Devriendt},
  {Dubois}, {Kashino}, {Paltani}, \& {Pichon}}]{Davidzon19}
{Davidzon}, I., {Laigle}, C., {Capak}, P.~L., {et~al.} 2019, \mnras, 489, 4817,
  \dodoi{10.1093/mnras/stz2486}

\bibitem[{{Davidzon} {et~al.}(2022){Davidzon}, {Jegatheesan}, {Ilbert}, {de la
  Torre}, {Leslie}, {Laigle}, {Hemmati}, {Masters}, {Blanquez-Sese},
  {Kauffmann}, {Magdis}, {Ma{\l}ek}, {McCracken}, {Mobasher}, {Moneti},
  {Sanders}, {Shuntov}, {Toft}, \& {Weaver}}]{Davidzon22}
{Davidzon}, I., {Jegatheesan}, K., {Ilbert}, O., {et~al.} 2022, \aap, 665, A34,
  \dodoi{10.1051/0004-6361/202243249}

\bibitem[{{de la Vega} {et~al.}(2025){de la Vega}, {Kassin}, {Pacifici},
  {Charlot}, {Curtis-Lake}, {Chevallard}, {Heckman}, {Koekemoer}, \&
  {Wang}}]{delavega25}
{de la Vega}, A., {Kassin}, S.~A., {Pacifici}, C., {et~al.} 2025, arXiv
  e-prints, arXiv:2501.06297.
\newblock \doarXiv{2501.06297}

\bibitem[{{Deshmukh} {et~al.}(2018){Deshmukh}, {Caputi}, {Ashby}, {Cowley},
  {McCracken}, {Fynbo}, {Le F{\`e}vre}, {Milvang-Jensen}, \&
  {Ilbert}}]{Deshmukh18}
{Deshmukh}, S., {Caputi}, K.~I., {Ashby}, M.~L.~N., {et~al.} 2018, \apj, 864,
  166, \dodoi{10.3847/1538-4357/aad9f5}

\bibitem[{{D'Eugenio} {et~al.}(2020){D'Eugenio}, {Daddi}, {Gobat},
  {Strazzullo}, {Lustig}, {Delvecchio}, {Jin}, {Puglisi}, {Calabr{\'o}},
  {Mancini}, {Dickinson}, {Cimatti}, \& {Onodera}}]{DEugenio20}
{D'Eugenio}, C., {Daddi}, E., {Gobat}, R., {et~al.} 2020, \apjl, 892, L2,
  \dodoi{10.3847/2041-8213/ab7a96}

\bibitem[{{D'Eugenio} {et~al.}(2024){D'Eugenio}, {Cameron}, {Scholtz},
  {Carniani}, {Willott}, {Curtis-Lake}, {Bunker}, {Parlanti}, {Maiolino},
  {Willmer}, {Jakobsen}, {Robertson}, {Johnson}, {Tacchella}, {Cargile},
  {Rawle}, {Arribas}, {Chevallard}, {Curti}, {Egami}, {Eisenstein}, {Kumari},
  {Looser}, {Rieke}, {Rodr{\'\i}guez Del Pino}, {Saxena}, {{\"U}bler},
  {Venturi}, {Witstok}, {Baker}, {Bhatawdekar}, {Bonaventura}, {Boyett},
  {Charlot}, {Danhaive}, {Hainline}, {Hausen}, {Helton}, {Ji}, {Ji}, {Jones},
  {Joud{\v{z}}balis}, {Maseda}, {P{\'e}rez-Gonz{\'a}lez}, {Perna},
  {Pusk{\'a}s}, {Shivaei}, {Silcock}, {Simmonds}, {Smit}, {Sun}, {Villanueva},
  {Williams}, \& {Zhu}}]{DEugenio24}
{D'Eugenio}, F., {Cameron}, A.~J., {Scholtz}, J., {et~al.} 2024, arXiv
  e-prints, arXiv:2404.06531, \dodoi{10.48550/arXiv.2404.06531}

\bibitem[{{Dunlop} {et~al.}(2021){Dunlop}, {Abraham}, {Ashby}, {Bagley},
  {Best}, {Bongiorno}, {Bouwens}, {Bowler}, {Brammer}, {Bremer}, {Calabro'},
  {Carnall}, {Castellano}, {Cirasuolo}, {Conselice}, {Cullen}, {Dave}, {Dayal},
  {Dekel}, {Dickinson}, {Duncan}, {Elbaz}, {Ellis}, {Ferguson}, {Ferrara},
  {Finkelstein}, {Fontana}, {Furlanetto}, {Fynbo}, {Gallerani}, {Gardner},
  {Giavalisco}, {Grazian}, {Grogin}, {Harikane}, {Hopkins}, {Ilbert},
  {Illingworth}, {Juneau}, {Jung}, {Kartaltepe}, {Kassin}, {Kauffmann},
  {Khochfar}, {Kirkpatrick}, {Kocevski}, {Koekemoer}, {Labbe}, {Laporte},
  {Larson}, {Lucas}, {Magee}, {Mason}, {McCracken}, {McLeod}, {McLure},
  {Merlin}, {Mesinger}, {Milvang-Jensen}, {Newman}, {Oesch}, {Ouchi},
  {Pacifici}, {Papovich}, {Peacock}, {Peeples}, {Pentericci}, {Perez-Gonzalez},
  {Pirzkal}, {Pope}, {Pye}, {Reddy}, {Robertson}, {Salvato}, {Santini},
  {Schaerer}, {Shapley}, {Simons}, {Smit}, {Smith}, {Snyder}, {Somerville},
  {Stanway}, {Stefanon}, {Tasca}, {Tikkanen}, {Tresse}, {Trump}, {Whitaker},
  {Wilkins}, {Wright}, {Wyithe}, {van Dokkum}, \& {van der Werf}}]{Dunlop21}
{Dunlop}, J.~S., {Abraham}, R.~G., {Ashby}, M. L.~N., {et~al.} 2021, {PRIMER:
  Public Release IMaging for Extragalactic Research}, JWST Proposal. Cycle 1,
  ID. \#1837

\bibitem[{{Eisenstein} {et~al.}(2023{\natexlab{a}}){Eisenstein}, {Willott},
  {Alberts}, {Arribas}, {Bonaventura}, {Bunker}, {Cameron}, {Carniani},
  {Charlot}, {Curtis-Lake}, {D'Eugenio}, {Endsley}, {Ferruit}, {Giardino},
  {Hainline}, {Hausen}, {Jakobsen}, {Johnson}, {Maiolino}, {Rieke}, {Rieke},
  {Rix}, {Robertson}, {Stark}, {Tacchella}, {Williams}, {Willmer}, {Baker},
  {Baum}, {Bhatawdekar}, {Boyett}, {Chen}, {Chevallard}, {Circosta}, {Curti},
  {Danhaive}, {DeCoursey}, {de Graaff}, {Dressler}, {Egami}, {Helton},
  {Hviding}, {Ji}, {Jones}, {Kumari}, {L{\"u}tzgendorf}, {Laseter}, {Looser},
  {Lyu}, {Maseda}, {Nelson}, {Parlanti}, {Perna}, {Pusk{\'a}s}, {Rawle},
  {Rodr{\'\i}guez Del Pino}, {Sandles}, {Saxena}, {Scholtz}, {Sharpe},
  {Shivaei}, {Silcock}, {Simmonds}, {Skarbinski}, {Smit}, {Stone}, {Suess},
  {Sun}, {Tang}, {Topping}, {{\"U}bler}, {Villanueva}, {Wallace}, {Whitler},
  {Witstok}, \& {Woodrum}}]{Eisenstein23a}
{Eisenstein}, D.~J., {Willott}, C., {Alberts}, S., {et~al.} 2023{\natexlab{a}},
  arXiv e-prints, arXiv:2306.02465, \dodoi{10.48550/arXiv.2306.02465}

\bibitem[{{Eisenstein} {et~al.}(2023{\natexlab{b}}){Eisenstein}, {Johnson},
  {Robertson}, {Tacchella}, {Hainline}, {Jakobsen}, {Maiolino}, {Bonaventura},
  {Bunker}, {Cameron}, {Cargile}, {Curtis-Lake}, {Hausen}, {Pusk{\'a}s},
  {Rieke}, {Sun}, {Willmer}, {Willott}, {Alberts}, {Arribas}, {Baker}, {Baum},
  {Bhatawdekar}, {Carniani}, {Charlot}, {Chen}, {Chevallard}, {Curti},
  {DeCoursey}, {D'Eugenio}, {de Graaff}, {Egami}, {Helton}, {Ji}, {Jones},
  {Kumari}, {L{\"u}tzgendorf}, {Laseter}, {Looser}, {Lyu}, {Maseda}, {Nelson},
  {Parlanti}, {Rauscher}, {Rawle}, {Rieke}, {Rix}, {Rujopakarn}, {Sandles},
  {Saxena}, {Scholtz}, {Sharpe}, {Shivaei}, {Simmonds}, {Smit}, {Topping},
  {{\"U}bler}, {Venturi}, {Williams}, {Witstok}, \& {Woodrum}}]{Eisenstein23b}
{Eisenstein}, D.~J., {Johnson}, B.~D., {Robertson}, B., {et~al.}
  2023{\natexlab{b}}, arXiv e-prints, arXiv:2310.12340,
  \dodoi{10.48550/arXiv.2310.12340}

\bibitem[{{Eldridge} {et~al.}(2017){Eldridge}, {Stanway}, {Xiao}, {McClelland},
  {Taylor}, {Ng}, {Greis}, \& {Bray}}]{Eldridge17}
{Eldridge}, J.~J., {Stanway}, E.~R., {Xiao}, L., {et~al.} 2017, \pasa, 34,
  e058, \dodoi{10.1017/pasa.2017.51}

\bibitem[{{Faisst} {et~al.}(2019){Faisst}, {Prakash}, {Capak}, \&
  {Lee}}]{Faisst19}
{Faisst}, A.~L., {Prakash}, A., {Capak}, P.~L., \& {Lee}, B. 2019, \apjl, 881,
  L9, \dodoi{10.3847/2041-8213/ab3581}

\bibitem[{{Ferland} {et~al.}(2017){Ferland}, {Chatzikos}, {Guzm{\'a}n},
  {Lykins}, {van Hoof}, {Williams}, {Abel}, {Badnell}, {Keenan}, {Porter}, \&
  {Stancil}}]{Ferland17}
{Ferland}, G.~J., {Chatzikos}, M., {Guzm{\'a}n}, F., {et~al.} 2017, \rmxaa, 53,
  385, \dodoi{10.48550/arXiv.1705.10877}

\bibitem[{{Feroz} {et~al.}(2009){Feroz}, {Hobson}, \& {Bridges}}]{Feroz09}
{Feroz}, F., {Hobson}, M.~P., \& {Bridges}, M. 2009, \mnras, 398, 1601,
  \dodoi{10.1111/j.1365-2966.2009.14548.x}

\bibitem[{{Finkelstein} {et~al.}(2022){Finkelstein}, {Bagley}, {Arrabal Haro},
  {Dickinson}, {Ferguson}, {Kartaltepe}, {Papovich}, {Burgarella}, {Kocevski},
  {Huertas-Company}, {Iyer}, {Koekemoer}, {Larson}, {P{\'e}rez-Gonz{\'a}lez},
  {Rose}, {Tacchella}, {Wilkins}, {Chworowsky}, {Medrano}, {Morales},
  {Somerville}, {Yung}, {Fontana}, {Giavalisco}, {Grazian}, {Grogin}, {Kewley},
  {Kirkpatrick}, {Kurczynski}, {Lotz}, {Pentericci}, {Pirzkal}, {Ravindranath},
  {Ryan}, {Trump}, {Yang}, {Almaini}, {Amor{\'\i}n}, {Annunziatella},
  {Backhaus}, {Barro}, {Behroozi}, {Bell}, {Bhatawdekar}, {Bisigello}, {Bromm},
  {Buat}, {Buitrago}, {Calabr{\`o}}, {Casey}, {Castellano}, {Ch{\'a}vez Ortiz},
  {Ciesla}, {Cleri}, {Cohen}, {Cole}, {Cooke}, {Cooper}, {Cooray}, {Costantin},
  {Cox}, {Croton}, {Daddi}, {Dav{\'e}}, {de La Vega}, {Dekel}, {Elbaz},
  {Estrada-Carpenter}, {Faber}, {Fern{\'a}ndez}, {Finkelstein}, {Freundlich},
  {Fujimoto}, {Garc{\'\i}a-Argum{\'a}nez}, {Gardner}, {Gawiser},
  {G{\'o}mez-Guijarro}, {Guo}, {Hamblin}, {Hamilton}, {Hathi}, {Holwerda},
  {Hirschmann}, {Hutchison}, {Jaskot}, {Jha}, {Jogee}, {Juneau}, {Jung},
  {Kassin}, {Bail}, {Leung}, {Lucas}, {Magnelli}, {Mantha}, {Matharu},
  {McGrath}, {McIntosh}, {Merlin}, {Mobasher}, {Newman}, {Nicholls}, {Pandya},
  {Rafelski}, {Ronayne}, {Santini}, {Seill{\'e}}, {Shah}, {Shen}, {Simons},
  {Snyder}, {Stanway}, {Straughn}, {Teplitz}, {Vanderhoof}, {Vega-Ferrero},
  {Wang}, {Weiner}, {Willmer}, {Wuyts}, {Zavala}, \& {Ceers
  Team}}]{Finkelstein22}
{Finkelstein}, S.~L., {Bagley}, M.~B., {Arrabal Haro}, P., {et~al.} 2022,
  \apjl, 940, L55, \dodoi{10.3847/2041-8213/ac966e}

\bibitem[{{Finkelstein} {et~al.}(2023){Finkelstein}, {Bagley}, {Ferguson},
  {Wilkins}, {Kartaltepe}, {Papovich}, {Yung}, {Haro}, {Behroozi}, {Dickinson},
  {Kocevski}, {Koekemoer}, {Larson}, {Le Bail}, {Morales},
  {P{\'e}rez-Gonz{\'a}lez}, {Burgarella}, {Dav{\'e}}, {Hirschmann},
  {Somerville}, {Wuyts}, {Bromm}, {Casey}, {Fontana}, {Fujimoto}, {Gardner},
  {Giavalisco}, {Grazian}, {Grogin}, {Hathi}, {Hutchison}, {Jha}, {Jogee},
  {Kewley}, {Kirkpatrick}, {Long}, {Lotz}, {Pentericci}, {Pierel}, {Pirzkal},
  {Ravindranath}, {Ryan}, {Trump}, {Yang}, {Bhatawdekar}, {Bisigello}, {Buat},
  {Calabr{\`o}}, {Castellano}, {Cleri}, {Cooper}, {Croton}, {Daddi}, {Dekel},
  {Elbaz}, {Franco}, {Gawiser}, {Holwerda}, {Huertas-Company}, {Jaskot},
  {Leung}, {Lucas}, {Mobasher}, {Pandya}, {Tacchella}, {Weiner}, \&
  {Zavala}}]{Finkelstein23}
{Finkelstein}, S.~L., {Bagley}, M.~B., {Ferguson}, H.~C., {et~al.} 2023, \apjl,
  946, L13, \dodoi{10.3847/2041-8213/acade4}

\bibitem[{{Finkelstein} {et~al.}(2025){Finkelstein}, {Bagley}, {Arrabal Haro},
  {Dickinson}, {Ferguson}, {Kartaltepe}, {Kocevski}, {Koekemoer}, {Lotz},
  {Papovich}, {Perez-Gonzalez}, {Pirzkal}, {Somerville}, {Trump}, {Yang},
  {Yung}, {Fontana}, {Grazian}, {Grogin}, {Kewley}, {Kirkpatrick}, {Larson},
  {Pentericci}, {Ravindranath}, {Wilkins}, {Almaini}, {Amorin}, {Barro},
  {Bhatawdekar}, {Bisigello}, {Brooks}, {Buitrago}, {Calabro}, {Castellano},
  {Cheng}, {Cleri}, {Cole}, {Cooper}, {Cooper}, {Costantin}, {Cox}, {Croton},
  {Daddi}, {Davis}, {Dekel}, {Elbaz}, {Fernandez}, {Fujimoto}, {Gandolfi},
  {Gardner}, {Gawiser}, {Giavalisco}, {Gomez-Guijarro}, {Guo}, {Gupta},
  {Hathi}, {Harish}, {Henry}, {Hirschmann}, {Hu}, {Hutchison}, {Iyer},
  {Jaskot}, {Jha}, {Jung}, {Kokorev}, {Kurczynski}, {Leung}, {Llerena}, {Long},
  {Lucas}, {Lu}, {McGrath}, {McIntosh}, {Merlin}, {Morales}, {Napolitano},
  {Pacucci}, {Pandya}, {Rafelski}, {Rodighiero}, {Rose}, {Santini}, {Seille},
  {Simons}, {Shen}, {Straughn}, {Tacchella}, {Vanderhoof}, {Vega-Ferrero},
  {Weiner}, {Willmer}, {Zhu}, {Bell}, {Wuyts}, {Holwerda}, {Wang}, {Wang}, \&
  {Zavala}}]{Finkelstein25}
{Finkelstein}, S.~L., {Bagley}, M.~B., {Arrabal Haro}, P., {et~al.} 2025, arXiv
  e-prints, arXiv:2501.04085, \dodoi{10.48550/arXiv.2501.04085}

\bibitem[{{Forrest} {et~al.}(2020){Forrest}, {Annunziatella}, {Wilson},
  {Marchesini}, {Muzzin}, {Cooper}, {Marsan}, {McConachie}, {Chan}, {Gomez},
  {Kado-Fong}, {L Barbera}, {Labb{\'e}}, {Lange-Vagle}, {Nantais}, {Nonino},
  {Pe{\~n}a}, {Saracco}, {Stefanon}, \& {van der Burg}}]{Forrest20}
{Forrest}, B., {Annunziatella}, M., {Wilson}, G., {et~al.} 2020, \apjl, 890,
  L1, \dodoi{10.3847/2041-8213/ab5b9f}

\bibitem[{{Gallazzi} {et~al.}(2014){Gallazzi}, {Bell}, {Zibetti}, {Brinchmann},
  \& {Kelson}}]{Gallazzi14}
{Gallazzi}, A., {Bell}, E.~F., {Zibetti}, S., {Brinchmann}, J., \& {Kelson},
  D.~D. 2014, \apj, 788, 72, \dodoi{10.1088/0004-637X/788/1/72}

\bibitem[{{Gardner} {et~al.}(2023){Gardner}, {Mather}, {Abbott}, {Abell},
  {Abernathy}, {Abney}, {Abraham}, {Abraham}, {Abul-Huda}, {Acton}, {Adams},
  {Adams}, {Adler}, {Adriaensen}, {Aguilar}, {Ahmed}, {Ahmed}, {Ahmed},
  {Albat}, {Albert}, {Alberts}, {Aldridge}, {Allen}, {Allen}, {Altenburg},
  {Altunc}, {Alvarez}, {{\'A}lvarez-M{\'a}rquez}, {Alves de Oliveira},
  {Ambrose}, {Anandakrishnan}, {Andersen}, {Anderson}, {Anderson}, {Anderson},
  {Anderson}, {Aprea}, {Archer}, {Arenberg}, {Argyriou}, {Arribas}, {Artigau},
  {Arvai}, {Atcheson}, {Atkinson}, {Averbukh}, {Aymergen}, {Bacinski},
  {Baggett}, {Bagnasco}, {Baker}, {Balzano}, {Banks}, {Baran}, {Barker},
  {Barrett}, {Barringer}, {Barto}, {Bast}, {Baudoz}, {Baum}, {Beatty},
  {Beaulieu}, {Bechtold}, {Beck}, {Beddard}, {Beichman}, {Bellagama}, {Bely},
  {Berger}, {Bergeron}, {Bernier}, {Bertch}, {Beskow}, {Betz}, {Biagetti},
  {Birkmann}, {Bjorklund}, {Blackwood}, {Blazek}, {Blossfeld}, {Bluth},
  {Boccaletti}, {Boegner}, {Bohlin}, {Boia}, {B{\"o}ker}, {Bonaventura},
  {Bond}, {Bosley}, {Boucarut}, {Bouchet}, {Bouwman}, {Bower}, {Bowers},
  {Bowers}, {Boyce}, {Boyer}, {Boyer}, {Boyer}, {Boyer}, {Bradley}, {Brady},
  {Brandl}, {Brannen}, {Breda}, {Bremmer}, {Brennan}, {Bresnahan}, {Bright},
  {Broiles}, {Bromenschenkel}, {Brooks}, {Brooks}, {Brown}, {Brown}, {Brown},
  {Bruce}, {Bryson}, {Bujanda}, {Bullock}, {Bunker}, {Bureo}, {Burt}, {Bush},
  {Bushouse}, {Bussman}, {Cabaud}, {Cale}, {Calhoon}, {Calvani}, {Canipe},
  {Caputo}, {Cara}, {Carey}, {Case}, {Cesari}, {Cetorelli}, {Chance},
  {Chandler}, {Chaney}, {Chapman}, {Charlot}, {Chayer}, {Cheezum}, {Chen},
  {Chen}, {Cherinka}, {Chichester}, {Chilton}, {Chittiraibalan}, {Clampin},
  {Clark}, {Clark}, {Clark}, {Claybrooks}, {Cleveland}, {Cohen}, {Cohen},
  {Col{\'o}n}, {Coleman}, {Colina}, {Comber}, {Comeau}, {Comer}, {Conde Reis},
  {Connolly}, {Conroy}, {Contos}, {Contreras}, {Cook}, {Cooper}, {Cooper},
  {Correia}, {Correnti}, {Cossou}, {Costanza}, {Coulais}, {Cox}, {Coyle},
  {Cracraft}, {Crew}, {Curtis}, {Cusveller}, {Da Costa Maciel}, {Dailey},
  {Daugeron}, {Davidson}, {Davies}, {Davis}, {Davis}, {Day}, {de Chambure}, {de
  Jong}, {De Marchi}, {Dean}, {Decker}, {Delisa}, {Dell}, \&
  {Dellagatta}}]{Gardner23}
{Gardner}, J.~P., {Mather}, J.~C., {Abbott}, R., {et~al.} 2023, \pasp, 135,
  068001, \dodoi{10.1088/1538-3873/acd1b5}

\bibitem[{{Geach}(2012)}]{Geach12}
{Geach}, J.~E. 2012, \mnras, 419, 2633,
  \dodoi{10.1111/j.1365-2966.2011.19913.x}

\bibitem[{{Gehrels}(1986)}]{Gehrels86}
{Gehrels}, N. 1986, \apj, 303, 336, \dodoi{10.1086/164079}

\bibitem[{{Giavalisco} {et~al.}(2004){Giavalisco}, {Ferguson}, {Koekemoer},
  {Dickinson}, {Alexander}, {Bauer}, {Bergeron}, {Biagetti}, {Brandt},
  {Casertano}, {Cesarsky}, {Chatzichristou}, {Conselice}, {Cristiani}, {Da
  Costa}, {Dahlen}, {de Mello}, {Eisenhardt}, {Erben}, {Fall}, {Fassnacht},
  {Fosbury}, {Fruchter}, {Gardner}, {Grogin}, {Hook}, {Hornschemeier}, {Idzi},
  {Jogee}, {Kretchmer}, {Laidler}, {Lee}, {Livio}, {Lucas}, {Madau},
  {Mobasher}, {Moustakas}, {Nonino}, {Padovani}, {Papovich}, {Park},
  {Ravindranath}, {Renzini}, {Richardson}, {Riess}, {Rosati}, {Schirmer},
  {Schreier}, {Somerville}, {Spinrad}, {Stern}, {Stiavelli}, {Strolger},
  {Urry}, {Vandame}, {Williams}, \& {Wolf}}]{Giavalisco04}
{Giavalisco}, M., {Ferguson}, H.~C., {Koekemoer}, A.~M., {et~al.} 2004, \apjl,
  600, L93, \dodoi{10.1086/379232}

\bibitem[{{Girelli} {et~al.}(2019){Girelli}, {Bolzonella}, \&
  {Cimatti}}]{Girelli19}
{Girelli}, G., {Bolzonella}, M., \& {Cimatti}, A. 2019, \aap, 632, A80,
  \dodoi{10.1051/0004-6361/201834547}

\bibitem[{{Glazebrook} {et~al.}(2017){Glazebrook}, {Schreiber}, {Labb{\'e}},
  {Nanayakkara}, {Kacprzak}, {Oesch}, {Papovich}, {Spitler}, {Straatman},
  {Tran}, \& {Yuan}}]{Glazebrook17}
{Glazebrook}, K., {Schreiber}, C., {Labb{\'e}}, I., {et~al.} 2017, \nat, 544,
  71, \dodoi{10.1038/nature21680}

\bibitem[{{Gould} {et~al.}(2023){Gould}, {Brammer}, {Valentino}, {Whitaker},
  {Weaver}, {Lagos}, {Rizzo}, {Franco}, {Hsieh}, {Ilbert}, {Jin}, {Magdis},
  {McCracken}, {Mobasher}, {Shuntov}, {Steinhardt}, {Strait}, \&
  {Toft}}]{Gould23}
{Gould}, K. M.~L., {Brammer}, G., {Valentino}, F., {et~al.} 2023, \aj, 165,
  248, \dodoi{10.3847/1538-3881/accadc}

\bibitem[{{Grogin} {et~al.}(2011){Grogin}, {Kocevski}, {Faber}, {Ferguson},
  {Koekemoer}, {Riess}, {Acquaviva}, {Alexander}, {Almaini}, {Ashby}, {Barden},
  {Bell}, {Bournaud}, {Brown}, {Caputi}, {Casertano}, {Cassata}, {Castellano},
  {Challis}, {Chary}, {Cheung}, {Cirasuolo}, {Conselice}, {Roshan Cooray},
  {Croton}, {Daddi}, {Dahlen}, {Dav{\'e}}, {de Mello}, {Dekel}, {Dickinson},
  {Dolch}, {Donley}, {Dunlop}, {Dutton}, {Elbaz}, {Fazio}, {Filippenko},
  {Finkelstein}, {Fontana}, {Gardner}, {Garnavich}, {Gawiser}, {Giavalisco},
  {Grazian}, {Guo}, {Hathi}, {H{\"a}ussler}, {Hopkins}, {Huang}, {Huang},
  {Jha}, {Kartaltepe}, {Kirshner}, {Koo}, {Lai}, {Lee}, {Li}, {Lotz}, {Lucas},
  {Madau}, {McCarthy}, {McGrath}, {McIntosh}, {McLure}, {Mobasher},
  {Moustakas}, {Mozena}, {Nandra}, {Newman}, {Niemi}, {Noeske}, {Papovich},
  {Pentericci}, {Pope}, {Primack}, {Rajan}, {Ravindranath}, {Reddy}, {Renzini},
  {Rix}, {Robaina}, {Rodney}, {Rosario}, {Rosati}, {Salimbeni}, {Scarlata},
  {Siana}, {Simard}, {Smidt}, {Somerville}, {Spinrad}, {Straughn}, {Strolger},
  {Telford}, {Teplitz}, {Trump}, {van der Wel}, {Villforth}, {Wechsler},
  {Weiner}, {Wiklind}, {Wild}, {Wilson}, {Wuyts}, {Yan}, \& {Yun}}]{Grogin11}
{Grogin}, N.~A., {Kocevski}, D.~D., {Faber}, S.~M., {et~al.} 2011, \apjs, 197,
  35, \dodoi{10.1088/0067-0049/197/2/35}

\bibitem[{{Hainline} {et~al.}(2024){Hainline}, {Johnson}, {Robertson},
  {Tacchella}, {Helton}, {Sun}, {Eisenstein}, {Simmonds}, {Topping}, {Whitler},
  {Willmer}, {Rieke}, {Suess}, {Hviding}, {Cameron}, {Alberts}, {Baker},
  {Baum}, {Bhatawdekar}, {Bonaventura}, {Boyett}, {Bunker}, {Carniani},
  {Charlot}, {Chevallard}, {Chen}, {Curti}, {Curtis-Lake}, {D'Eugenio},
  {Egami}, {Endsley}, {Hausen}, {Ji}, {Looser}, {Lyu}, {Maiolino}, {Nelson},
  {Pusk{\'a}s}, {Rawle}, {Sandles}, {Saxena}, {Smit}, {Stark}, {Williams},
  {Willott}, \& {Witstok}}]{Hainline24}
{Hainline}, K.~N., {Johnson}, B.~D., {Robertson}, B., {et~al.} 2024, \apj, 964,
  71, \dodoi{10.3847/1538-4357/ad1ee4}

\bibitem[{Harris {et~al.}(2020)Harris, Millman, van~der Walt, Gommers,
  Virtanen, Cournapeau, Wieser, Taylor, Berg, Smith, Kern, Picus, Hoyer, van
  Kerkwijk, Brett, Haldane, del R{\'{i}}o, Wiebe, Peterson,
  G{\'{e}}rard-Marchant, Sheppard, Reddy, Weckesser, Abbasi, Gohlke, \&
  Oliphant}]{harris2020array}
Harris, C.~R., Millman, K.~J., van~der Walt, S.~J., {et~al.} 2020, Nature, 585,
  357, \dodoi{10.1038/s41586-020-2649-2}

\bibitem[{{Hemmati} {et~al.}(2019){Hemmati}, {Capak}, {Pourrahmani}, {Nayyeri},
  {Stern}, {Mobasher}, {Darvish}, {Davidzon}, {Ilbert}, {Masters}, \&
  {Shahidi}}]{Hemmati19}
{Hemmati}, S., {Capak}, P., {Pourrahmani}, M., {et~al.} 2019, \apjl, 881, L14,
  \dodoi{10.3847/2041-8213/ab3418}

\bibitem[{{Huertas-Company} \& {Lanusse}(2023)}]{HuertasCompany23}
{Huertas-Company}, M., \& {Lanusse}, F. 2023, \pasa, 40, e001,
  \dodoi{10.1017/pasa.2022.55}

\bibitem[{Hunter(2007)}]{Hunter:2007}
Hunter, J.~D. 2007, Computing in Science \& Engineering, 9, 90,
  \dodoi{10.1109/MCSE.2007.55}

\bibitem[{{Jin} {et~al.}(2024){Jin}, {Sillassen}, {Magdis}, {Brinch},
  {Shuntov}, {Brammer}, {Gobat}, {Valentino}, {Carnall}, {Lee}, {Vijayan},
  {Gillman}, {Kokorev}, {Le Bail}, {Greve}, {Gullberg}, {Gould}, \&
  {Toft}}]{Jin24}
{Jin}, S., {Sillassen}, N.~B., {Magdis}, G.~E., {et~al.} 2024, \aap, 683, L4,
  \dodoi{10.1051/0004-6361/202348540}

\bibitem[{{Kodra} {et~al.}(2023){Kodra}, {Andrews}, {Newman}, {Finkelstein},
  {Fontana}, {Hathi}, {Salvato}, {Wiklind}, {Wuyts}, {Broussard}, {Chartab},
  {Conselice}, {Cooper}, {Dekel}, {Dickinson}, {Ferguson}, {Gawiser}, {Grogin},
  {Iyer}, {Kartaltepe}, {Kassin}, {Koekemoer}, {Koo}, {Lucas}, {Mantha},
  {McIntosh}, {Mobasher}, {Pacifici}, {P{\'e}rez-Gonz{\'a}lez}, \&
  {Santini}}]{Kodra23}
{Kodra}, D., {Andrews}, B.~H., {Newman}, J.~A., {et~al.} 2023, \apj, 942, 36,
  \dodoi{10.3847/1538-4357/ac9f12}

\bibitem[{{Koekemoer} {et~al.}(2011){Koekemoer}, {Faber}, {Ferguson}, {Grogin},
  {Kocevski}, {Koo}, {Lai}, {Lotz}, {Lucas}, {McGrath}, {Ogaz}, {Rajan},
  {Riess}, {Rodney}, {Strolger}, {Casertano}, {Castellano}, {Dahlen},
  {Dickinson}, {Dolch}, {Fontana}, {Giavalisco}, {Grazian}, {Guo}, {Hathi},
  {Huang}, {van der Wel}, {Yan}, {Acquaviva}, {Alexander}, {Almaini}, {Ashby},
  {Barden}, {Bell}, {Bournaud}, {Brown}, {Caputi}, {Cassata}, {Challis},
  {Chary}, {Cheung}, {Cirasuolo}, {Conselice}, {Roshan Cooray}, {Croton},
  {Daddi}, {Dav{\'e}}, {de Mello}, {de Ravel}, {Dekel}, {Donley}, {Dunlop},
  {Dutton}, {Elbaz}, {Fazio}, {Filippenko}, {Finkelstein}, {Frazer}, {Gardner},
  {Garnavich}, {Gawiser}, {Gruetzbauch}, {Hartley}, {H{\"a}ussler},
  {Herrington}, {Hopkins}, {Huang}, {Jha}, {Johnson}, {Kartaltepe},
  {Khostovan}, {Kirshner}, {Lani}, {Lee}, {Li}, {Madau}, {McCarthy},
  {McIntosh}, {McLure}, {McPartland}, {Mobasher}, {Moreira}, {Mortlock},
  {Moustakas}, {Mozena}, {Nandra}, {Newman}, {Nielsen}, {Niemi}, {Noeske},
  {Papovich}, {Pentericci}, {Pope}, {Primack}, {Ravindranath}, {Reddy},
  {Renzini}, {Rix}, {Robaina}, {Rosario}, {Rosati}, {Salimbeni}, {Scarlata},
  {Siana}, {Simard}, {Smidt}, {Snyder}, {Somerville}, {Spinrad}, {Straughn},
  {Telford}, {Teplitz}, {Trump}, {Vargas}, {Villforth}, {Wagner}, {Wandro},
  {Wechsler}, {Weiner}, {Wiklind}, {Wild}, {Wilson}, {Wuyts}, \&
  {Yun}}]{Koekemoer11}
{Koekemoer}, A.~M., {Faber}, S.~M., {Ferguson}, H.~C., {et~al.} 2011, \apjs,
  197, 36, \dodoi{10.1088/0067-0049/197/2/36}

\bibitem[{{Kohonen}(2001)}]{Kohonen01}
{Kohonen}, T. 2001, {Self-Organizing Maps}

\bibitem[{{La Torre} {et~al.}(2024){La Torre}, {Sajina}, {Goulding},
  {Marchesini}, {Bezanson}, {Pearl}, \& {Sodr{\'e}}}]{LaTorre24}
{La Torre}, V., {Sajina}, A., {Goulding}, A.~D., {et~al.} 2024, \aj, 167, 261,
  \dodoi{10.3847/1538-3881/ad3821}

\bibitem[{{Labb{\'e}} {et~al.}(2005){Labb{\'e}}, {Huang}, {Franx}, {Rudnick},
  {Barmby}, {Daddi}, {van Dokkum}, {Fazio}, {F{\"o}rster Schreiber},
  {Moorwood}, {Rix}, {R{\"o}ttgering}, {Trujillo}, \& {van der Werf}}]{Labbe05}
{Labb{\'e}}, I., {Huang}, J., {Franx}, M., {et~al.} 2005, \apjl, 624, L81,
  \dodoi{10.1086/430700}

\bibitem[{{Lagos} {et~al.}(2024){Lagos}, {Bravo}, {Tobar}, {Obreschkow},
  {Power}, {Robotham}, {Proctor}, {Hansen}, {Chandro-G{\'o}mez}, \&
  {Carrivick}}]{Lagos24}
{Lagos}, C. d.~P., {Bravo}, M., {Tobar}, R., {et~al.} 2024, \mnras, 531, 3551,
  \dodoi{10.1093/mnras/stae1024}

\bibitem[{{Leja} {et~al.}(2019){Leja}, {Johnson}, {Conroy}, {van Dokkum},
  {Speagle}, {Brammer}, {Momcheva}, {Skelton}, {Whitaker}, {Franx}, \&
  {Nelson}}]{Leja19}
{Leja}, J., {Johnson}, B.~D., {Conroy}, C., {et~al.} 2019, \apj, 877, 140,
  \dodoi{10.3847/1538-4357/ab1d5a}

\bibitem[{{Long} {et~al.}(2023){Long}, {Casey}, {del P. Lagos}, {Lambrides},
  {Zavala}, {Champagne}, {Cooper}, \& {Cooray}}]{Long23}
{Long}, A.~S., {Casey}, C.~M., {del P. Lagos}, C., {et~al.} 2023, \apj, 953,
  11, \dodoi{10.3847/1538-4357/acddde}

\bibitem[{{Long} {et~al.}(2024){Long}, {Antwi-Danso}, {Lambrides}, {Lovell},
  {de la Vega}, {Valentino}, {Zavala}, {Casey}, {Wilkins}, {Yung}, {Arrabal
  Haro}, {Bagley}, {Bisigello}, {Chworowsky}, {Cooper}, {Cooper}, {Cooray},
  {Croton}, {Dickinson}, {Finkelstein}, {Franco}, {Gould}, {Hirschmann},
  {Hutchison}, {Kartaltepe}, {Kocevski}, {Koekemoer}, {Lucas}, {McKinney},
  {Nere}, {Papovich}, {P{\'e}rez-Gonz{\'a}lez}, {Pirzkal}, \&
  {Santini}}]{Long24}
{Long}, A.~S., {Antwi-Danso}, J., {Lambrides}, E.~L., {et~al.} 2024, \apj, 970,
  68, \dodoi{10.3847/1538-4357/ad4cea}

\bibitem[{{Lovell} {et~al.}(2021){Lovell}, {Vijayan}, {Thomas}, {Wilkins},
  {Barnes}, {Irodotou}, \& {Roper}}]{Lovell21}
{Lovell}, C.~C., {Vijayan}, A.~P., {Thomas}, P.~A., {et~al.} 2021, \mnras, 500,
  2127, \dodoi{10.1093/mnras/staa3360}

\bibitem[{{Lovell} {et~al.}(2023){Lovell}, {Roper}, {Vijayan}, {Seeyave},
  {Irodotou}, {Wilkins}, {Conselice}, {Fortuni}, {Kuusisto}, {Merlin},
  {Santini}, \& {Thomas}}]{Lovell23}
{Lovell}, C.~C., {Roper}, W., {Vijayan}, A.~P., {et~al.} 2023, \mnras, 525,
  5520, \dodoi{10.1093/mnras/stad2550}

\bibitem[{{Lustig} {et~al.}(2023){Lustig}, {Strazzullo}, {Remus}, {D'Eugenio},
  {Daddi}, {Burkert}, {De Lucia}, {Delvecchio}, {Dolag}, {Fontanot}, {Gobat},
  {Mohr}, {Onodera}, {Pannella}, \& {Pillepich}}]{Lustig23}
{Lustig}, P., {Strazzullo}, V., {Remus}, R.-S., {et~al.} 2023, \mnras, 518,
  5953, \dodoi{10.1093/mnras/stac3450}

\bibitem[{{Martis} {et~al.}(2019){Martis}, {Marchesini}, {Muzzin}, {Stefanon},
  {Brammer}, {da Cunha}, {Sajina}, \& {Labbe}}]{Martis19}
{Martis}, N.~S., {Marchesini}, D.~M., {Muzzin}, A., {et~al.} 2019, \apj, 882,
  65, \dodoi{10.3847/1538-4357/ab32f1}

\bibitem[{{Masters} {et~al.}(2015){Masters}, {Capak}, {Stern}, {Ilbert},
  {Salvato}, {Schmidt}, {Longo}, {Rhodes}, {Paltani}, {Mobasher}, {Hoekstra},
  {Hildebrandt}, {Coupon}, {Steinhardt}, {Speagle}, {Faisst}, {Kalinich},
  {Brodwin}, {Brescia}, \& {Cavuoti}}]{Masters15}
{Masters}, D., {Capak}, P., {Stern}, D., {et~al.} 2015, \apj, 813, 53,
  \dodoi{10.1088/0004-637X/813/1/53}

\bibitem[{{McInnes} {et~al.}(2018){McInnes}, {Healy}, \&
  {Melville}}]{McInnes18}
{McInnes}, L., {Healy}, J., \& {Melville}, J. 2018, arXiv e-prints,
  arXiv:1802.03426, \dodoi{10.48550/arXiv.1802.03426}

\bibitem[{{McKinney} {et~al.}(2023){McKinney}, {Finnerty}, {Casey}, {Franco},
  {Long}, {Fujimoto}, {Zavala}, {Cooper}, {Akins}, {Pope}, {Armus}, {Soifer},
  {Larson}, {Matthews}, {Melbourne}, \& {Cushing}}]{McKinney23}
{McKinney}, J., {Finnerty}, L., {Casey}, C.~M., {et~al.} 2023, \apjl, 946, L39,
  \dodoi{10.3847/2041-8213/acc322}

\bibitem[{{Merlin} {et~al.}(2018){Merlin}, {Fontana}, {Castellano}, {Santini},
  {Torelli}, {Boutsia}, {Wang}, {Grazian}, {Pentericci}, {Schreiber}, {Ciesla},
  {McLure}, {Derriere}, {Dunlop}, \& {Elbaz}}]{Merlin18}
{Merlin}, E., {Fontana}, A., {Castellano}, M., {et~al.} 2018, \mnras, 473,
  2098, \dodoi{10.1093/mnras/stx2385}

\bibitem[{{Merlin} {et~al.}(2019){Merlin}, {Fortuni}, {Torelli}, {Santini},
  {Castellano}, {Fontana}, {Grazian}, {Pentericci}, {Pilo}, \&
  {Schmidt}}]{Merlin19}
{Merlin}, E., {Fortuni}, F., {Torelli}, M., {et~al.} 2019, \mnras, 490, 3309,
  \dodoi{10.1093/mnras/stz2615}

\bibitem[{{Nanayakkara} {et~al.}(2024){Nanayakkara}, {Glazebrook}, {Jacobs},
  {Kawinwanichakij}, {Schreiber}, {Brammer}, {Esdaile}, {Kacprzak}, {Labbe},
  {Lagos}, {Marchesini}, {Marsan}, {Oesch}, {Papovich}, {Remus}, \&
  {Tran}}]{Nanayakkara24}
{Nanayakkara}, T., {Glazebrook}, K., {Jacobs}, C., {et~al.} 2024, Scientific
  Reports, 14, 3724, \dodoi{10.1038/s41598-024-52585-4}

\bibitem[{{Nayyeri} {et~al.}(2014){Nayyeri}, {Mobasher}, {Hemmati}, {De
  Barros}, {Ferguson}, {Wiklind}, {Dahlen}, {Dickinson}, {Giavalisco},
  {Fontana}, {Ashby}, {Barro}, {Guo}, {Hathi}, {Kassin}, {Koekemoer},
  {Willner}, {Dunlop}, {Paris}, \& {Targett}}]{Nayyeri14}
{Nayyeri}, H., {Mobasher}, B., {Hemmati}, S., {et~al.} 2014, \apj, 794, 68,
  \dodoi{10.1088/0004-637X/794/1/68}

\bibitem[{{Oke} \& {Gunn}(1983)}]{OkeGunn83}
{Oke}, J.~B., \& {Gunn}, J.~E. 1983, \apj, 266, 713, \dodoi{10.1086/160817}

\bibitem[{{Pacifici} {et~al.}(2016){Pacifici}, {Kassin}, {Weiner}, {Holden},
  {Gardner}, {Faber}, {Ferguson}, {Koo}, {Primack}, {Bell}, {Dekel}, {Gawiser},
  {Giavalisco}, {Rafelski}, {Simons}, {Barro}, {Croton}, {Dav{\'e}}, {Fontana},
  {Grogin}, {Koekemoer}, {Lee}, {Salmon}, {Somerville}, \&
  {Behroozi}}]{Pacifici16}
{Pacifici}, C., {Kassin}, S.~A., {Weiner}, B.~J., {et~al.} 2016, \apj, 832, 79,
  \dodoi{10.3847/0004-637X/832/1/79}

\bibitem[{{Pacifici} {et~al.}(2023){Pacifici}, {Iyer}, {Mobasher}, {da Cunha},
  {Acquaviva}, {Burgarella}, {Calistro Rivera}, {Carnall}, {Chang}, {Chartab},
  {Cooke}, {Fairhurst}, {Kartaltepe}, {Leja}, {Ma{\l}ek}, {Salmon}, {Torelli},
  {Vidal-Garc{\'\i}a}, {Boquien}, {Brammer}, {Brown}, {Capak}, {Chevallard},
  {Circosta}, {Croton}, {Davidzon}, {Dickinson}, {Duncan}, {Faber}, {Ferguson},
  {Fontana}, {Guo}, {Haeussler}, {Hemmati}, {Jafariyazani}, {Kassin}, {Larson},
  {Lee}, {Mantha}, {Marchi}, {Nayyeri}, {Newman}, {Pandya}, {Pforr}, {Reddy},
  {Sanders}, {Shah}, {Shahidi}, {Stevans}, {Triani}, {Tyler}, {Vanderhoof}, {de
  la Vega}, {Wang}, \& {Weston}}]{Pacifici23}
{Pacifici}, C., {Iyer}, K.~G., {Mobasher}, B., {et~al.} 2023, \apj, 944, 141,
  \dodoi{10.3847/1538-4357/acacff}

\bibitem[{{Park} {et~al.}(2023){Park}, {Belli}, {Conroy}, {Tacchella}, {Leja},
  {Cutler}, {Johnson}, {Nelson}, \& {Emami}}]{Park23}
{Park}, M., {Belli}, S., {Conroy}, C., {et~al.} 2023, \apj, 953, 119,
  \dodoi{10.3847/1538-4357/acd54a}

\bibitem[{Pedregosa {et~al.}(2011)Pedregosa, Varoquaux, Gramfort, Michel,
  Thirion, Grisel, Blondel, Prettenhofer, Weiss, Dubourg, Vanderplas, Passos,
  Cournapeau, Brucher, Perrot, \& Duchesnay}]{scikit-learn}
Pedregosa, F., Varoquaux, G., Gramfort, A., {et~al.} 2011, Journal of Machine
  Learning Research, 12, 2825

\bibitem[{{Planck Collaboration} {et~al.}(2016){Planck Collaboration}, {Ade},
  {Aghanim}, {Arnaud}, {Ashdown}, {Aumont}, {Baccigalupi}, {Banday},
  {Barreiro}, {Bartlett}, {Bartolo}, {Battaner}, {Battye}, {Benabed},
  {Beno{\^\i}t}, {Benoit-L{\'e}vy}, {Bernard}, {Bersanelli}, {Bielewicz},
  {Bock}, {Bonaldi}, {Bonavera}, {Bond}, {Borrill}, {Bouchet}, {Boulanger},
  {Bucher}, {Burigana}, {Butler}, {Calabrese}, {Cardoso}, {Catalano},
  {Challinor}, {Chamballu}, {Chary}, {Chiang}, {Chluba}, {Christensen},
  {Church}, {Clements}, {Colombi}, {Colombo}, {Combet}, {Coulais}, {Crill},
  {Curto}, {Cuttaia}, {Danese}, {Davies}, {Davis}, {de Bernardis}, {de Rosa},
  {de Zotti}, {Delabrouille}, {D{\'e}sert}, {Di Valentino}, {Dickinson},
  {Diego}, {Dolag}, {Dole}, {Donzelli}, {Dor{\'e}}, {Douspis}, {Ducout},
  {Dunkley}, {Dupac}, {Efstathiou}, {Elsner}, {En{\ss}lin}, {Eriksen},
  {Farhang}, {Fergusson}, {Finelli}, {Forni}, {Frailis}, {Fraisse},
  {Franceschi}, {Frejsel}, {Galeotta}, {Galli}, {Ganga}, {Gauthier}, {Gerbino},
  {Ghosh}, {Giard}, {Giraud-H{\'e}raud}, {Giusarma}, {Gjerl{\o}w},
  {Gonz{\'a}lez-Nuevo}, {G{\'o}rski}, {Gratton}, {Gregorio}, {Gruppuso},
  {Gudmundsson}, {Hamann}, {Hansen}, {Hanson}, {Harrison}, {Helou},
  {Henrot-Versill{\'e}}, {Hern{\'a}ndez-Monteagudo}, {Herranz}, {Hildebrandt},
  {Hivon}, {Hobson}, {Holmes}, {Hornstrup}, {Hovest}, {Huang}, {Huffenberger},
  {Hurier}, {Jaffe}, {Jaffe}, {Jones}, {Juvela}, {Keih{\"a}nen}, {Keskitalo},
  {Kisner}, {Kneissl}, {Knoche}, {Knox}, {Kunz}, {Kurki-Suonio}, {Lagache},
  {L{\"a}hteenm{\"a}ki}, {Lamarre}, {Lasenby}, {Lattanzi}, {Lawrence}, {Leahy},
  {Leonardi}, {Lesgourgues}, {Levrier}, {Lewis}, {Liguori}, {Lilje},
  {Linden-V{\o}rnle}, {L{\'o}pez-Caniego}, {Lubin}, {Mac{\'\i}as-P{\'e}rez},
  {Maggio}, {Maino}, {Mandolesi}, {Mangilli}, {Marchini}, {Maris}, {Martin},
  {Martinelli}, {Mart{\'\i}nez-Gonz{\'a}lez}, {Masi}, {Matarrese}, {McGehee},
  {Meinhold}, {Melchiorri}, {Melin}, {Mendes}, {Mennella}, {Migliaccio},
  {Millea}, {Mitra}, {Miville-Desch{\^e}nes}, {Moneti}, {Montier}, {Morgante},
  {Mortlock}, {Moss}, {Munshi}, {Murphy}, {Naselsky}, {Nati}, {Natoli},
  {Netterfield}, {N{\o}rgaard-Nielsen}, {Noviello}, {Novikov}, {Novikov},
  {Oxborrow}, {Paci}, {Pagano}, {Pajot}, {Paladini}, {Paoletti}, {Partridge},
  {Pasian}, {Patanchon}, {Pearson}, {Perdereau}, {Perotto}, {Perrotta},
  {Pettorino}, {Piacentini}, {Piat}, {Pierpaoli}, {Pietrobon}, {Plaszczynski},
  {Pointecouteau}, {Polenta}, {Popa}, {Pratt}, {Pr{\'e}zeau}, {Prunet},
  {Puget}, {Rachen}, {Reach}, {Rebolo}, {Reinecke}, {Remazeilles}, {Renault},
  {Renzi}, {Ristorcelli}, {Rocha}, {Rosset}, {Rossetti}, {Roudier},
  {Rouill{\'e} d'Orfeuil}, {Rowan-Robinson}, {Rubi{\~n}o-Mart{\'\i}n},
  {Rusholme}, {Said}, {Salvatelli}, {Salvati}, {Sandri}, {Santos},
  {Savelainen}, {Savini}, {Scott}, {Seiffert}, {Serra}, {Shellard}, {Spencer},
  {Spinelli}, {Stolyarov}, {Stompor}, {Sudiwala}, {Sunyaev}, {Sutton},
  {Suur-Uski}, {Sygnet}, {Tauber}, {Terenzi}, {Toffolatti}, {Tomasi},
  {Tristram}, {Trombetti}, {Tucci}, {Tuovinen}, {T{\"u}rler}, {Umana},
  {Valenziano}, {Valiviita}, {Van Tent}, {Vielva}, {Villa}, {Wade}, {Wandelt},
  {Wehus}, {White}, {White}, {Wilkinson}, {Yvon}, {Zacchei}, \&
  {Zonca}}]{Planck16}
{Planck Collaboration}, {Ade}, P.~A.~R., {Aghanim}, N., {et~al.} 2016, \aap,
  594, A13, \dodoi{10.1051/0004-6361/201525830}

\bibitem[{{Rieke} {et~al.}(2023){Rieke}, {Robertson}, {Tacchella}, {Hainline},
  {Johnson}, {Hausen}, {Ji}, {Willmer}, {Eisenstein}, {Pusk{\'a}s}, {Alberts},
  {Arribas}, {Baker}, {Baum}, {Bhatawdekar}, {Bonaventura}, {Boyett}, {Bunker},
  {Cameron}, {Carniani}, {Charlot}, {Chevallard}, {Chen}, {Curti},
  {Curtis-Lake}, {Danhaive}, {DeCoursey}, {Dressler}, {Egami}, {Endsley},
  {Helton}, {Hviding}, {Kumari}, {Looser}, {Lyu}, {Maiolino}, {Maseda},
  {Nelson}, {Rieke}, {Rix}, {Sandles}, {Saxena}, {Sharpe}, {Shivaei},
  {Skarbinski}, {Smit}, {Stark}, {Stone}, {Suess}, {Sun}, {Topping},
  {{\"U}bler}, {Villanueva}, {Wallace}, {Williams}, {Willott}, {Whitler},
  {Witstok}, \& {Woodrum}}]{Rieke23}
{Rieke}, M.~J., {Robertson}, B., {Tacchella}, S., {et~al.} 2023, \apjs, 269,
  16, \dodoi{10.3847/1538-4365/acf44d}

\bibitem[{{Russell} {et~al.}(2024){Russell}, {Dobric}, {Adams}, {Conselice},
  {Austin}, {Harvey}, {Trussler}, {Ferreira}, {Westcott}, {Harris},
  {Windhorst}, {Coe}, {Cohen}, {Driver}, {Frye}, {Grogin}, {Hathi}, {Jansen},
  {Koekemoer}, {Marshall}, {Ortiz III}, {Pirzkal}, {Robotham}, {Ryan Jr},
  {Summers}, {D'Silva}, {Willmer}, \& {Yan}}]{Russell24}
{Russell}, T.~A., {Dobric}, N., {Adams}, N.~J., {et~al.} 2024, arXiv e-prints,
  arXiv:2412.11861.
\newblock \doarXiv{2412.11861}

\bibitem[{{Salim} {et~al.}(2018){Salim}, {Boquien}, \& {Lee}}]{Salim18}
{Salim}, S., {Boquien}, M., \& {Lee}, J.~C. 2018, \apj, 859, 11,
  \dodoi{10.3847/1538-4357/aabf3c}

\bibitem[{{Sanjaripour} {et~al.}(2024){Sanjaripour}, {Hemmati}, {Mobasher},
  {Canalizo}, {Barish}, {Shivaei}, {Coil}, {Chartab}, {Jafariyazani}, {Reddy},
  \& {Azadi}}]{Sanjaripour24}
{Sanjaripour}, S., {Hemmati}, S., {Mobasher}, B., {et~al.} 2024, arXiv
  e-prints, arXiv:2410.07354, \dodoi{10.48550/arXiv.2410.07354}

\bibitem[{{Santini} {et~al.}(2019){Santini}, {Merlin}, {Fontana}, {Magnelli},
  {Paris}, {Castellano}, {Grazian}, {Pentericci}, {Pilo}, \&
  {Torelli}}]{Santini19}
{Santini}, P., {Merlin}, E., {Fontana}, A., {et~al.} 2019, \mnras, 486, 560,
  \dodoi{10.1093/mnras/stz801}

\bibitem[{{Schaye} {et~al.}(2015){Schaye}, {Crain}, {Bower}, {Furlong},
  {Schaller}, {Theuns}, {Dalla Vecchia}, {Frenk}, {McCarthy}, {Helly},
  {Jenkins}, {Rosas-Guevara}, {White}, {Baes}, {Booth}, {Camps}, {Navarro},
  {Qu}, {Rahmati}, {Sawala}, {Thomas}, \& {Trayford}}]{Schaye15}
{Schaye}, J., {Crain}, R.~A., {Bower}, R.~G., {et~al.} 2015, \mnras, 446, 521,
  \dodoi{10.1093/mnras/stu2058}

\bibitem[{{Schreiber} {et~al.}(2018){Schreiber}, {Glazebrook}, {Nanayakkara},
  {Kacprzak}, {Labb{\'e}}, {Oesch}, {Yuan}, {Tran}, {Papovich}, {Spitler}, \&
  {Straatman}}]{Schreiber18}
{Schreiber}, C., {Glazebrook}, K., {Nanayakkara}, T., {et~al.} 2018, \aap, 618,
  A85, \dodoi{10.1051/0004-6361/201833070}

\bibitem[{{Setton} {et~al.}(2024){Setton}, {Khullar}, {Miller}, {Bezanson},
  {Greene}, {Suess}, {Whitaker}, {Antwi-Danso}, {Atek}, {Brammer}, {Cutler},
  {Dayal}, {Feldmann}, {Fujimoto}, {Furtak}, {Glazebrook}, {Goulding},
  {Kokorev}, {Labbe}, {Leja}, {Ma}, {Marchesini}, {Nanayakkara}, {Pan},
  {Price}, {Siegel}, {Shipley}, {Weaver}, {van Dokkum}, {Wang}, \&
  {Williams}}]{Setton24}
{Setton}, D.~J., {Khullar}, G., {Miller}, T.~B., {et~al.} 2024, \apj, 974, 145,
  \dodoi{10.3847/1538-4357/ad6a18}

\bibitem[{{Shah} {et~al.}(2024){Shah}, {Lemaux}, {Forrest}, {Cucciati}, {Hung},
  {Staab}, {Hathi}, {Lubin}, {Gal}, {Shen}, {Zamorani}, {Giddings}, {Bardelli},
  {Pasqua Cassara}, {Cassata}, {Contini}, {Golden-Marx}, {Guaita}, {Gururajan},
  {Koekemoer}, {McLeod}, {Tasca}, {Tresse}, {Vergani}, \& {Zucca}}]{Shah24}
{Shah}, E.~A., {Lemaux}, B., {Forrest}, B., {et~al.} 2024, \mnras, 529, 873,
  \dodoi{10.1093/mnras/stae519}

\bibitem[{{Shahidi} {et~al.}(2020){Shahidi}, {Mobasher}, {Nayyeri}, {Hemmati},
  {Wiklind}, {Chartab}, {Dickinson}, {Finkelstein}, {Pacifici}, {Papovich},
  {Ferguson}, {Fontana}, {Giavalisco}, {Koekemoer}, {Newman}, {Sattari}, \&
  {Somerville}}]{Shahidi20}
{Shahidi}, A., {Mobasher}, B., {Nayyeri}, H., {et~al.} 2020, \apj, 897, 44,
  \dodoi{10.3847/1538-4357/ab96c5}

\bibitem[{{Shen} {et~al.}(2024){Shen}, {Zhuang}, {Li}, {Burgasser}, {Fan},
  {Greene}, {Narayan}, {Shapley}, {Sun}, {Wang}, \& {Yang}}]{Shen24}
{Shen}, Y., {Zhuang}, M.-Y., {Li}, J., {et~al.} 2024, arXiv e-prints,
  arXiv:2408.12713, \dodoi{10.48550/arXiv.2408.12713}

\bibitem[{{Siegel} {et~al.}(2024){Siegel}, {Setton}, {Greene}, {Suess},
  {Whitaker}, {Bezanson}, {Leja}, {Furtak}, {Cutler}, {de Graaff}, {Feldmann},
  {Khullar}, {Labb{\'e}}, {Marchesini}, {Miller}, {Nanayakkara}, {Pan},
  {Price}, {Treiber}, {van Dokkum}, {Wang}, \& {Weaver}}]{Siegel24}
{Siegel}, J., {Setton}, D., {Greene}, J., {et~al.} 2024, arXiv e-prints,
  arXiv:2409.11457, \dodoi{10.48550/arXiv.2409.11457}

\bibitem[{{Simet} {et~al.}(2021){Simet}, {Chartab}, {Lu}, \&
  {Mobasher}}]{Simet21}
{Simet}, M., {Chartab}, N., {Lu}, Y., \& {Mobasher}, B. 2021, \apj, 908, 47,
  \dodoi{10.3847/1538-4357/abd179}

\bibitem[{{Skelton} {et~al.}(2014){Skelton}, {Whitaker}, {Momcheva}, {Brammer},
  {van Dokkum}, {Labb{\'e}}, {Franx}, {van der Wel}, {Bezanson}, {Da Cunha},
  {Fumagalli}, {F{\"o}rster Schreiber}, {Kriek}, {Leja}, {Lundgren}, {Magee},
  {Marchesini}, {Maseda}, {Nelson}, {Oesch}, {Pacifici}, {Patel}, {Price},
  {Rix}, {Tal}, {Wake}, \& {Wuyts}}]{Skelton14}
{Skelton}, R.~E., {Whitaker}, K.~E., {Momcheva}, I.~G., {et~al.} 2014, \apjs,
  214, 24, \dodoi{10.1088/0067-0049/214/2/24}

\bibitem[{{Somerville} {et~al.}(2015){Somerville}, {Popping}, \&
  {Trager}}]{Somerville15}
{Somerville}, R.~S., {Popping}, G., \& {Trager}, S.~C. 2015, \mnras, 453, 4337,
  \dodoi{10.1093/mnras/stv1877}

\bibitem[{{Somerville} {et~al.}(2021){Somerville}, {Olsen}, {Yung}, {Pacifici},
  {Ferguson}, {Behroozi}, {Osborne}, {Wechsler}, {Pandya}, {Faber}, {Primack},
  \& {Dekel}}]{Somerville21}
{Somerville}, R.~S., {Olsen}, C., {Yung}, L.~Y.~A., {et~al.} 2021, \mnras, 502,
  4858, \dodoi{10.1093/mnras/stab231}

\bibitem[{{Stanway} \& {Eldridge}(2018)}]{StanwayEldridge18}
{Stanway}, E.~R., \& {Eldridge}, J.~J. 2018, \mnras, 479, 75,
  \dodoi{10.1093/mnras/sty1353}

\bibitem[{{Steinhardt} {et~al.}(2016){Steinhardt}, {Capak}, {Masters}, \&
  {Speagle}}]{Steinhardt16}
{Steinhardt}, C.~L., {Capak}, P., {Masters}, D., \& {Speagle}, J.~S. 2016,
  \apj, 824, 21, \dodoi{10.3847/0004-637X/824/1/21}

\bibitem[{{Steinhardt} {et~al.}(2020){Steinhardt}, {Weaver}, {Maxfield},
  {Davidzon}, {Faisst}, {Masters}, {Schemel}, \& {Toft}}]{Steinhardt20}
{Steinhardt}, C.~L., {Weaver}, J.~R., {Maxfield}, J., {et~al.} 2020, \apj, 891,
  136, \dodoi{10.3847/1538-4357/ab76be}

\bibitem[{{Stevans} {et~al.}(2021){Stevans}, {Finkelstein}, {Kawinwanichakij},
  {Wold}, {Papovich}, {Somerville}, {Yung}, {Sherman}, {Ciardullo}, {Dav{\'e}},
  {Florez}, {Gronwall}, \& {Jogee}}]{Stevans21}
{Stevans}, M.~L., {Finkelstein}, S.~L., {Kawinwanichakij}, L., {et~al.} 2021,
  \apj, 921, 58, \dodoi{10.3847/1538-4357/ac0cf6}

\bibitem[{{Straatman} {et~al.}(2016){Straatman}, {Spitler}, {Quadri},
  {Labb{\'e}}, {Glazebrook}, {Persson}, {Papovich}, {Tran}, {Brammer},
  {Cowley}, {Tomczak}, {Nanayakkara}, {Alcorn}, {Allen}, {Broussard}, {van
  Dokkum}, {Forrest}, {van Houdt}, {Kacprzak}, {Kawinwanichakij}, {Kelson},
  {Lee}, {McCarthy}, {Mehrtens}, {Monson}, {Murphy}, {Rees}, {Tilvi}, \&
  {Whitaker}}]{Straatman16}
{Straatman}, C. M.~S., {Spitler}, L.~R., {Quadri}, R.~F., {et~al.} 2016, \apj,
  830, 51, \dodoi{10.3847/0004-637X/830/1/51}

\bibitem[{{Toft} {et~al.}(2014){Toft}, {Smol{\v{c}}i{\'c}}, {Magnelli},
  {Karim}, {Zirm}, {Michalowski}, {Capak}, {Sheth}, {Schawinski}, {Krogager},
  {Wuyts}, {Sanders}, {Man}, {Lutz}, {Staguhn}, {Berta}, {Mccracken}, {Krpan},
  \& {Riechers}}]{Toft14}
{Toft}, S., {Smol{\v{c}}i{\'c}}, V., {Magnelli}, B., {et~al.} 2014, \apj, 782,
  68, \dodoi{10.1088/0004-637X/782/2/68}

\bibitem[{{Valentino} {et~al.}(2020){Valentino}, {Tanaka}, {Davidzon}, {Toft},
  {G{\'o}mez-Guijarro}, {Stockmann}, {Onodera}, {Brammer}, {Ceverino},
  {Faisst}, {Gallazzi}, {Hayward}, {Ilbert}, {Kubo}, {Magdis}, {Selsing},
  {Shimakawa}, {Sparre}, {Steinhardt}, {Yabe}, \& {Zabl}}]{Valentino20}
{Valentino}, F., {Tanaka}, M., {Davidzon}, I., {et~al.} 2020, \apj, 889, 93,
  \dodoi{10.3847/1538-4357/ab64dc}

\bibitem[{{Valentino} {et~al.}(2023){Valentino}, {Brammer}, {Gould}, {Kokorev},
  {Fujimoto}, {Jespersen}, {Vijayan}, {Weaver}, {Ito}, {Tanaka}, {Ilbert},
  {Magdis}, {Whitaker}, {Faisst}, {Gallazzi}, {Gillman}, {Gim{\'e}nez-Arteaga},
  {G{\'o}mez-Guijarro}, {Kubo}, {Heintz}, {Hirschmann}, {Oesch}, {Onodera},
  {Rizzo}, {Lee}, {Strait}, \& {Toft}}]{Valentino23}
{Valentino}, F., {Brammer}, G., {Gould}, K. M.~L., {et~al.} 2023, \apj, 947,
  20, \dodoi{10.3847/1538-4357/acbefa}

\bibitem[{{van der Maaten} \& {Hinton}(2008)}]{vdM_Hinton08}
{van der Maaten}, L., \& {Hinton}, G. 2008, {J. Mach. Learn. Res.,}, 9

\bibitem[{{Vani} {et~al.}(2025){Vani}, {Ayromlou}, {Kauffmann}, \&
  {Springel}}]{Vani25}
{Vani}, A., {Ayromlou}, M., {Kauffmann}, G., \& {Springel}, V. 2025, \mnras,
  536, 777, \dodoi{10.1093/mnras/stae2625}

\bibitem[{{Vijayan} {et~al.}(2021){Vijayan}, {Lovell}, {Wilkins}, {Thomas},
  {Barnes}, {Irodotou}, {Kuusisto}, \& {Roper}}]{Vijayan21}
{Vijayan}, A.~P., {Lovell}, C.~C., {Wilkins}, S.~M., {et~al.} 2021, \mnras,
  501, 3289, \dodoi{10.1093/mnras/staa3715}

\bibitem[{Virtanen {et~al.}(2020)Virtanen, Gommers, Oliphant, Haberland, Reddy,
  Cournapeau, Burovski, Peterson, Weckesser, Bright, {van der Walt}, Brett,
  Wilson, Millman, Mayorov, Nelson, Jones, Kern, Larson, Carey, Polat, Feng,
  Moore, {VanderPlas}, Laxalde, Perktold, Cimrman, Henriksen, Quintero, Harris,
  Archibald, Ribeiro, Pedregosa, {van Mulbregt}, \& {SciPy 1.0
  Contributors}}]{2020SciPy-NMeth}
Virtanen, P., Gommers, R., Oliphant, T.~E., {et~al.} 2020, Nature Methods, 17,
  261, \dodoi{10.1038/s41592-019-0686-2}

\bibitem[{{Whitaker} {et~al.}(2011){Whitaker}, {Labb{\'e}}, {van Dokkum},
  {Brammer}, {Kriek}, {Marchesini}, {Quadri}, {Franx}, {Muzzin}, {Williams},
  {Bezanson}, {Illingworth}, {Lee}, {Lundgren}, {Nelson}, {Rudnick}, {Tal}, \&
  {Wake}}]{Whitaker11}
{Whitaker}, K.~E., {Labb{\'e}}, I., {van Dokkum}, P.~G., {et~al.} 2011, \apj,
  735, 86, \dodoi{10.1088/0004-637X/735/2/86}

\bibitem[{{Williams} {et~al.}(2018){Williams}, {Curtis-Lake}, {Hainline},
  {Chevallard}, {Robertson}, {Charlot}, {Endsley}, {Stark}, {Willmer},
  {Alberts}, {Amorin}, {Arribas}, {Baum}, {Bunker}, {Carniani}, {Crandall},
  {Egami}, {Eisenstein}, {Ferruit}, {Husemann}, {Maseda}, {Maiolino}, {Rawle},
  {Rieke}, {Smit}, {Tacchella}, \& {Willott}}]{Williams18}
{Williams}, C.~C., {Curtis-Lake}, E., {Hainline}, K.~N., {et~al.} 2018, \apjs,
  236, 33, \dodoi{10.3847/1538-4365/aabcbb}

\bibitem[{{Williams} {et~al.}(2023){Williams}, {Tacchella}, {Maseda},
  {Robertson}, {Johnson}, {Willott}, {Eisenstein}, {Willmer}, {Ji}, {Hainline},
  {Helton}, {Alberts}, {Baum}, {Bhatawdekar}, {Boyett}, {Bunker}, {Carniani},
  {Charlot}, {Chevallard}, {Curtis-Lake}, {de Graaff}, {Egami}, {Franx},
  {Kumari}, {Maiolino}, {Nelson}, {Rieke}, {Sandles}, {Shivaei}, {Simmonds},
  {Smit}, {Suess}, {Sun}, {{\"U}bler}, \& {Witstok}}]{Williams23}
{Williams}, C.~C., {Tacchella}, S., {Maseda}, M.~V., {et~al.} 2023, \apjs, 268,
  64, \dodoi{10.3847/1538-4365/acf130}

\bibitem[{{Williams} {et~al.}(2024){Williams}, {Oesch}, {Weibel}, {Brammer},
  {Cloonan}, {Whitaker}, {Barrufet}, {Bezanson}, {Bowler}, {Dayal}, {Franx},
  {Greene}, {Hutter}, {Ji}, {Labb{\'e}}, {Manning}, {Maseda}, \&
  {Xiao}}]{Williams24}
{Williams}, C.~C., {Oesch}, P.~A., {Weibel}, A., {et~al.} 2024, arXiv e-prints,
  arXiv:2410.01875, \dodoi{10.48550/arXiv.2410.01875}

\bibitem[{{Williams} {et~al.}(2009){Williams}, {Quadri}, {Franx}, {van Dokkum},
  \& {Labb{\'e}}}]{Williams09}
{Williams}, R.~J., {Quadri}, R.~F., {Franx}, M., {van Dokkum}, P., \&
  {Labb{\'e}}, I. 2009, \apj, 691, 1879, \dodoi{10.1088/0004-637X/691/2/1879}

\bibitem[{{Yung} {et~al.}(2019){Yung}, {Somerville}, {Finkelstein}, {Popping},
  \& {Dav{\'e}}}]{Yung19}
{Yung}, L.~Y.~A., {Somerville}, R.~S., {Finkelstein}, S.~L., {Popping}, G., \&
  {Dav{\'e}}, R. 2019, \mnras, 483, 2983, \dodoi{10.1093/mnras/sty3241}

\bibitem[{{Yung} {et~al.}(2022){Yung}, {Somerville}, {Ferguson}, {Finkelstein},
  {Gardner}, {Dav{\'e}}, {Bagley}, {Popping}, \& {Behroozi}}]{Yung22}
{Yung}, L.~Y.~A., {Somerville}, R.~S., {Ferguson}, H.~C., {et~al.} 2022,
  \mnras, 515, 5416, \dodoi{10.1093/mnras/stac2139}

\end{thebibliography}
\bibliographystyle{aasjournal}



\end{document}